\documentclass[12pt]{article}

\usepackage{amsmath}
\usepackage{amsthm}

\usepackage{amssymb}

\usepackage{epsfig}

\begin{document}

\newtheorem{theorem}{Theorem}[section]
\newtheorem{definition}{Definition}[section]
\newtheorem{lemma}{Lemma}[section]
\newtheorem{corollary}{Corollary}[section]
\newtheorem{proposition}{Proposition}[section]
\newtheorem*{lemma*}{Lemma}
\newtheorem{conjecture}{Conjecture}[section]
\newtheorem*{conjecture*}{Conjecture}
\newtheorem*{corollary*}{Corollary}

\newcommand{\half}{\frac{1}{2}}

\renewcommand{\theequation}{\thesection.\arabic{equation}}

\newcommand{\barray}{\setlength\arraycolsep{2pt} \begin{array}}
\newcommand{\earray}{\end{array}}
\newcommand{\dis}{\displaystyle}

\newcommand{\ope}{operator product expansion}

\newcommand{\NO}[2]{{:} #1 #2 {:} }
\newcommand{\NOthree}[3]{{:} #1 #2 #3 {:}}
\newcommand{\com}[2]{\left[#1,#2\right]}
\newcommand{\acom}[2]{\{#1,#2\}}

\newcommand{\dd}{\partial}

\newcommand{\ket}[1]{ \big|\, #1  \, \big>}

\newcommand{\fg}{\mathfrak{g}}
\newcommand{\fh}{\mathfrak{h}}
\newcommand{\fn}{\mathfrak{n}}
\newcommand{\ag}{\widehat{\mathfrak{g}}}
\newcommand{\twg}{\widetilde{\mathfrak{g}}}

\newcommand{\e}{\epsilon}

\newcommand{\ad}[1]{\text{ad}\,#1}

\newcommand{\ghb}{{b}}
\newcommand{\ghg}{{c}}

\newcommand{\str}{{\mathrm f}}

\newcommand{\nz}{{\mathbb{N}_0}}
\newcommand{\ZZ}{{\mathbb{Z}}}
\newcommand{\al}{\alpha}


\begin{titlepage}

\phantom{a}

\vfill

\begin{center}
{\LARGE \bf
Ramond sector of superconformal algebras via quantum reduction\\
}
\vspace{20pt}
{
{ \large \bf Boris Noyvert\footnote{
Work partially supported by Minerva Foundation, Germany. Grant No.~8466.
}
\\} 
\vspace{10pt}
{  e-mail: 
\hspace{-15pt}
\raisebox{-0.7pt}{
\includegraphics[height=8pt]{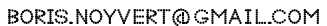}
}\\}
\vspace{3pt}
{ \small \it Department of Mathematics,}\\
{ \small \it  Weizmann Institute of Science,} \\
{ \small \it  76100, Rehovot,  Israel.}\\
}

\vspace{20pt}

\begin{abstract}

{\normalsize

Quantum hamiltonian reduction of affine superalgebras
is studied in the twisted case.
The Ramond sector of ``minimal'' superconformal W-algebras is
described in detail, the determinant formula
is obtained. Extensive list of examples includes
all the simple Lie superalgebras of rank up to 2.
The paper generalizes
the results of Kac and Wakimoto to the twisted case.
}

\end{abstract}

\end{center}

\vfill

\end{titlepage}

\tableofcontents



\section{Introduction}

\label{Introduction}

\setcounter{equation}{0}


Quantum hamiltonian reduction applied to affine superalgebras
leads to superconformal W-algebras, which are infinite dimensional
algebras with relations that are polynomial in the generators.
From the physical point of view the quantum hamiltonian reduction
is a procedure of BRST quantization of WZWN models with
constraints. Then the W-algebra is a symmetry algebra
of the constrained model.

The quantum reduction associates to every $\frac{1}{2}\mathbb{Z}$ gradation
on a Lie superalgebra $\fg$ with a non--zero
even invariant supersymmetric bilinear form
a BRST complex, the homology of which is the W-algebra.
Non--equivalent gradations on the same Lie superalgebra lead to different
W-algebras. Typical $\frac{1}{2}\mathbb{Z}$ gradations are those
generated by $sl(2)$ embeddings into $\fg$: under the action of
the $sl(2)$ subalgebra the algebra $\fg$ decouples to a sum of
$sl(2)$ eigenspaces with half--integer
eigenvalues.

The quantum hamiltonian reduction allows not only to construct
W-algebras, but also to describe their representation theory.
The characters and the determinant formula of  highest weight
representations of the underlying affine algebra are translated
to the characters and the determinant formula of the correspondent
W-algebra representations. Since the theory of Lie superalgebras
and their Kac--Moody affinizations is relatively well developed,
the quantum reduction becomes a strong tool for study
superconformal W-algebras and their representation theory.

The study of hamiltonian reduction of Lie superalgebras
has a long history. The classical reduction is known
since 1980's \cite{Drinfeld:1984qv}. The quantization
of the classical reduction is developed in
\cite{Bershadsky:1989mf, Feigin:1990qn, Feigin:1990pn, Bershadsky:1990bg}.
The first three papers discuss the quantum hamiltonian reduction
of $sl(N)$, based on the
principal $sl(2)$ embedding to $sl(N)$, which gives rise
to the so called $W_N$ algebras
\cite{Zamolodchikov:1985wn, Fateev:1988zh}.
The paper by Bershadsky \cite{Bershadsky:1990bg}
is on the quantum reduction corresponding to the non-principal $sl(2)$
embedding to $sl(3)$. The reduction results in the so called
Bershadsky--Polyakov algebra.
In this case the constraints on the WZWN model are of the second class,
and ``auxiliary fields'' (``neutral free superfermions'' in the terminology
of the present paper) have been introduced to describe the second class
constraints.

The quantum reduction procedure was
further developed in \cite{Frenkel:1992ju}: the representation theory
of the W-algebra was connected to the representation theory of
the underlying affine algebra, in particular characters and
fusion coefficients of modular invariant representations
of $W_N$ algebras were calculated.

The subject of quantum reduction was under intensive study in early 1990's,
see for example \cite{Feher:1992yx} and references therein. The constrained WZWN models on
Lie superalgebras were studied in \cite{Frappat:1992bs}.

The quantum reduction theory was developed for the case of an integral
gradation only, or for a half--integral gradation which can be reduced to the integral one.
However some Lie superalgebras
have only half-integral gradations
(including the simplest one $osp(1|2)$).
The breakthrough was achieved only in 2003
in the series of papers by Kac et al \cite{Kac talk,KRW,KW1}:
the quantum reduction was constructed for any Lie superalgebra
with a non--zero even invariant supersymmetric bilinear form.
The structure of the resulting W-algebra was described in detail.
W-algebras corresponding to minimal gradations
(``minimal'' W-algebras)
were constructed explicitly.
The representation theory of ``minimal'' W-algebras is developed
in \cite{KW1}: the determinant formula is obtained.

The untwisted case only is discussed in the papers \cite{KRW,KW1}. However
the twisted sectors (e.g.~the Ramond sector)
of superconformal W-algebras are of great importance
in physics. In the present paper we generalize the procedure of quantum reduction
to the twisted case. The modifications are described in detail. The determinant
formula for ``minimal'' W-algebras is calculated in the twisted case.

The paper is organized as follows. In Section~\ref{Gradation on a
Lie superalgebra} we introduce the framework: we recall
from \cite{KRW} the definitions of gradations
on Lie superalgebras, ``good'' gradations, minimal gradations. In
Section~\ref{Ingredients} we collect all the necessary information
on the main ingredients of the construction: affine vertex
algebra, superghost system, neutral free superfermion system.
Special attention is devoted to the twisted case. We recall the
main points of the general quantum reduction procedure in
Section~\ref{Quantum reduction}. The modifications due to twisted
case are explained. In Section~\ref{Minimal W-algebras} we
concentrate on the
``minimal''
W-algebras.
In Section~\ref{Determinant formulae for minimal W-algebras}
we state and prove the determinant formula for the Ramond sector
representations of the ``minimal'' W-algebra.
Section~\ref{Examples} contains a list of examples: quantum reduction
of Lie superalgebras of rank up to two is briefly discussed
and explicit determinant formulas are presented.
Section~\ref{Discussion} contains the discussion of results and
their comparison to the results of \cite{KW2}.
Appendix~\ref{Normal ordered product conventions}
fixes the normal ordered product conventions.

When we finished the derivation of the results of the present
paper, a work by Kac and Wakimoto \cite{KW2} appeared on the net.
They consider the same subject and obtain essentially the same
results as in our paper. However, since there is a conceptual difference in
some technical details (see Section~\ref{Discussion})
and in the presentation style, we decided to publish our paper.

The highest root of a Lie superalgebra is conventionally
normalized by $(\theta|\theta)=2$ in the current paper.
``$\mathbb{N}$'' is used for positive integers,
``$\nz$'' -- for non-negative integers.


\section{Gradation on a Lie superalgebra}

\label{Gradation on a Lie superalgebra}

\setcounter{equation}{0}


We start from a simple finite dimensional Lie superalgebra $\fg$
with a non-degenerate
even supersymmetric invariant bilinear form $(.|.)$.
The gradation of $\fg$ is the linear space decomposition
\begin{equation}   \label{gradation def}
\fg=\bigoplus_j \fg_j, \qquad
\text{such that }
\com{\fg_i}{\fg_j} \subset \fg_{i+j}.
\end{equation}
We say the gradation is generated by an element $x\in \fg$,
if the subspaces $\fg_j$ are eigenspaces of $\ad{x}$ with
eigenvalue $j$: $\com{x}{u}=j\,u\,$ for $\, u\in \fg_j$.

Fix an even element $x\in \fg$, such that it generates
a gradation in $\fg$ with half-integer eigenvalues:
$\fg=\bigoplus_{j\in\half\mathbb{Z}} \fg_j$.
Denote
\begin{equation}                 \label{g>}
  \fg_>=\bigoplus_{j>0}\fg_j,
\qquad
\fg_\le=\bigoplus_{j \le 0}\fg_j.
\end{equation}
An even element $f\in \fg_{-1}$
is called good if its
centralizer $\fg^f=\{u\in\fg {\,|\,} \com{f}{u}=0 \}$
lies in $\fg_\le$ ($\fg^f \subset \fg_\le$).
A gradation is called good if it is generated by an even
element $x\in\fg_0$ with half-integer eigenvalues and admits
a good element $f\in\fg_{-1}$.

Typical examples of good gradations are gradations
associated to the $sl(2)$ embeddings in the Lie superalgebra.
They are called Dynkin gradations and generated
by an element of an $sl(2)$ triple.
Even elements $f,x,e \in \fg$ form an $sl(2)$ triple, if they satisfy
the commutation relations:
\begin{equation}
  \com{x}{e}=e, \qquad
\com{x}{f}=-f, \qquad
\com{e}{f}=x.
\end{equation}
It is known from the $sl(2)$ representation theory that
the gradation generated by $x$ is a good gradation.
There are many good non--Dynkin gradations.
Good gradations of simple Lie algebras are classified in
\cite{EK}.

If one chooses $x\in\fh$, where $\fh$ is the Cartan subalgebra
of $\fg$, then the root elements of $\fg$ have a well defined
grading. The gradation (\ref{gradation def}) generates the root
system decomposition: $\Delta=\bigcup_{j\in\frac{1}{2}\mathbb{Z}}
\Delta_j$, where
\begin{equation}   \label{Delta_j}
\Delta_{j}=\{\alpha\in\Delta \,|\, \alpha(x)=j \}.
\end{equation}
Define $\Delta_>$ to be a set of roots
corresponding to $\fg_>$:
\begin{equation}          \label{Delta greater}
\Delta_>=\{\alpha\in\Delta \,|\, \alpha(x)>0\}=\bigcup_{j>0} \Delta_j \, .
\end{equation}

In this paper we focus on the so called minimal gradations \cite{KRW}.
Minimal gradation is a Dynkin gradation by $\ad{x}$
\begin{equation}
  \fg=\fg_{-1} \oplus \fg_{-1/2} \oplus \fg_{0} \oplus
\fg_{1/2} \oplus \fg_{1},
\end{equation}
such that $\fg_{-1}$ and $\fg_{1}$ are even one--dimensional
spaces, i.e.~$\fg_{-1}=\mathbb{C} f$ and $\fg_{1}=\mathbb{C} e$,
and $x=\com{e}{f}$.

Minimal gradations are obtained in the Lie algebra case
by choosing $sl(2)$ embedding corresponding to the highest
root $\theta$: $e=u_\theta, f=u_{-\theta}$, where
$u_\theta$ is the highest
root element of $\fg$.
In the Lie superalgebra case the construction is the same,
$\theta$ is chosen to be the highest root of one
of the simple subalgebras of the even part of $\fg$.

Next we want to define the affine vertex algebra $V_k(\fg)$
associated to the Lie superalgebra $\fg$. In order to proceed
with a BRST quantization we should also introduce
two sets of ghost fields: the superghost system and
the superfermion system.


\section{Ingredients}

\label{Ingredients}

\setcounter{equation}{0}


In this section we introduce the main ingredients of the construction:
affine vertex algebra, superghost system,
neutral superfermion system.
The section may be red and used independently from the other parts
of the paper.


\subsection{Affine vertex algebra}

\label{Affine vertex algebra}


Let $\fg$ be a simple finite dimensional Lie superalgebra
with an even nondegenerate supersymmetric invariant
bilinear form $(.|.)$. 
One associates a current $u(z)$ to every $u \in \fg$.
The collection of fields $\{u(z)\}_{u \in \fg}$ together
with a level $k \in \mathbb{C}$ satisfying the following \ope s
\begin{equation}                            \label{affine_ope}
  u(z)\,v(w)=\frac{k \, (u|v)}{(z-w)^2}
+\frac{\com{u}{v}(w)}{z-w}, \qquad u,v \in \fg
\end{equation}
is called the universal affine vertex algebra $V_k(\fg)$.

Fix 
a triangular decomposition
$\fg=\fn^-\oplus \fh \oplus \fn^+$.
The Weyl vector $\rho$
is defined with respect to a corresponding
set of positive roots:
\begin{equation}
 2 \, (\rho|\alpha_i)=
(\alpha_i|\alpha_i),
\qquad i=1,2,\ldots, {\text{rank}\,\fg},
\end{equation}
where $\alpha_i$ are simple roots of $\fg$.
The Weyl vector can be 
computed as
\begin{equation}
  \rho=\frac{1}{2} \sum_{\alpha \in \Delta_+}
(-1)^{p(\alpha)} \alpha,
\end{equation}
where the sum is over the set of positive roots
$\Delta_+$ and $p(\alpha)=0$ (respectively 1) for $\alpha$
even (respectively odd).

The dual Coxeter number $h^\lor$ is defined as
one half of the eigenvalue of the Casimir operator
in the adjoint representation. It can be calculated as
\begin{equation}
  h^\lor=(\rho|\theta)+\frac{1}{2}(\theta|\theta),
\end{equation}
where $\theta$ is the highest root. 

Let $\{u_i\}$ and $\{u^i\}$ be a pair of dual bases
of $\fg$, i.e.~$(u_i|u^j)=\delta_i^j$. The energy--momentum
field for the affine vertex algebra $V_k(\fg)$ is given by
the Sugawara construction:
\begin{equation}  \label{L g}
  L^\fg =
\frac{1}{2(k+h^\lor)}
\sum_i 
\NO{u^i}{u_i}
\end{equation}
(assuming $k \ne -h^\lor$). The central charge of the Virasoro
algebra generated by $L^\fg$ is
\begin{equation}
  c_\fg = \frac{k}{k+h^\lor}\, \text{sdim}\fg.
\end{equation}
The currents $u(z)$ are primary of conformal dimension 1 with respect to
$L^\fg (z)$.
The mode expansion of the affine currents is
\begin{equation}
  u(z)=\sum_{n\in \epsilon(u)+\mathbb{Z}}
u_n z^{-n-1},
\end{equation}
where $\epsilon(u)\in\mathbb{R}/ 
\mathbb{Z}$ is called the twisting of the field $u$. The \ope\
(\ref{affine_ope}) leads to the commutation relations for the
modes:
\begin{equation}    \label{affine_comm}
  \com{u_m}{v_n}=m\,k\,\delta_{m+n,0}\, (u|v)+
{\com{u}{v}}_{m+n} .
\end{equation}
The choice of $\epsilon(u)$
should be consistent with the structure of $\fg$:
\begin{equation}  \label{twisting consistence}
\epsilon(\com{u}{v})-\epsilon(u)-\epsilon(v) \in \mathbb{Z}.
\end{equation}
In particular $\epsilon(u)=0, \forall u\in\fg$
(untwisted case) is always allowed.
In this paper we will deal only with the case
$\epsilon(h)=0$ for all $h\in\fh$.
(Although the case $\epsilon(h)=1/2$ for some $h\in\fh$
is not forbidden.) In this case all root elements $u_\alpha,\,
\alpha \in \Delta$ have a well defined twisting.
Then there is a ${\text{rank}\,\fg}$
continuous parameter family of twistings, defined as following:
\begin{equation}     \label{general affine twisting}
\barray{l}
\e(h)=0,\  h\in\fh,\\
\e(u_\alpha) \text{ is any number in } \mathbb{R}/ \mathbb{Z},\
 \alpha \text{ - simple root},
\earray
\end{equation}
and the twistings for the basis elements corresponding to the
non-simple roots are defined by (\ref{twisting consistence}).

In the untwisted case ($\epsilon(u)=0,\ \forall u\in \fg$)
the modes $m$ and $n$ in the commutation relation (\ref{affine_comm})
are integer. Then one recognizes that it is the defining Lie bracket
of affine superalgebra $\ag$, the Kac--Moody affinization of $\fg$.
The affine superalgebra $\ag$ is defined as an infinite dimensional Lie superalgebra
$\ag=\fg[t,t^{-1}]\oplus \mathbb{C} K \oplus \mathbb{C} D$
with commutation relations
\begin{equation}
\begin{aligned}
\com{u t^m}{v t^n}&=\com{u}{v} t^{m+n}
+ (u,v) \, m \, \delta_{m+n,0}\,  K,\\
\com{D}{a t^m}&=m a t^m, \qquad \com{K}{\ag}=0,
\end{aligned}
\end{equation}
where $u,v \in \fg,\ m,n \in \mathbb{Z}$.
Denoting $u_n\equiv u t^n$ and choosing
$K=k\, I$ we return to the commutation relation (\ref{affine_comm}).
$D$ acts on $\ag$ as a minus zero mode of the Sugawara energy--momentum
field: $D \sim - L^\fg_0\,$.

The universal affine vertex algebra $V_k(\fg)$ written in terms of
field modes can be understood as a generalization of the affine algebra
$\ag$ to the case of arbitrary twisting.
We will denote it by $\twg$ and call it a {\em twisted loop algebra}.
In many cases the twisted loop algebra $\twg$ is isomorphic
to the untwisted one $\ag$. In the case
(\ref{general affine twisting})
the isomorphism $\ag \to \twg$ is given by
$u_{\alpha,n} \mapsto u_{\alpha,n+\e(u)},\,
h(\alpha)_n \mapsto h(\alpha)_n+k\, \e(u_\alpha)\, \delta_{n,0},\,
n\in\mathbb{Z}$, $h(\alpha)\in \fh$ is the Cartan element
associated to the root $\alpha$.

There are different choices of triangular decomposition
of $\twg$.
We choose a natural generalization of the triangular decomposition
in the untwisted case\footnote{
Another choice is implemented in \cite{KW2}
}:
\begin{equation}      \label{triangular decomposition}
\begin{gathered}
\hfill
\twg=\widetilde \fn^- \oplus \widetilde \fh
\oplus \widetilde \fn^+, \hfill\\
\begin{aligned}
\widetilde \fh&=\mathbb{C}[\{h_0 {\,|\,} h\in\fh, \e(h)=0\}
\cup \{K,D\} ],\\
\widetilde \fn^+&= 
\mathbb{C}[\{u_n {\,|\,} n>0, u \in \fg\}
\cup 
\{u_0{\,|\,} u\in \fn^+, \e(u)=0\}],\\
\widetilde \fn^- &= 
\mathbb{C}[\{u_n {\,|\,} n<0, u \in \fg\}
\cup \{u_0{\,|\,}
u \in \fn^-,  \e(u)=0\} ].
\end{aligned}
\end{gathered}
\end{equation}

In the case $\e(\fh)=0$
the set of positive roots $\widehat \Delta_+$ of $\twg$ is a disjoint union of\footnote{
We denote vectors in the root space by triples
$\widehat{\alpha}=
(\widehat\alpha(\fh),\widehat\alpha(K),\widehat\alpha(D))$.
}
\begin{equation}               \label{affine positive roots}
\begin{gathered}
\{(\alpha,0,m)\, |\, \alpha \in \Delta_- ,\, m>0,\, m\in \e(u_\alpha)+\mathbb{Z} \}, \\
\{(\alpha,0,m)\, |\, \alpha \in \Delta_+ ,\, m\ge 0,\, m\in \e(u_\alpha)+\mathbb{Z}\},
\{(0,0,m)\, |\, m>0,\, m\in\mathbb{Z} \},
\end{gathered}
\end{equation}
where the multiplicity of the last set is $r=\text{rank}\, \fg$.
There are $r +1$ simple roots.
The supersymmetric invariant bilinear form of $\fg$
is extended to $\twg$ in the standard way:
\begin{equation}
  \begin{array}{c}
    (u_m|v_n)=(u|v) \, \delta_{m+n,0},\\
    (D|u_m)=0, \qquad (K|u_m)=0, \\
    (D|D)=0=(K|K),  \qquad (D|K)=1,
  \end{array}
\qquad
\begin{array}{c}
u,v \in \fg,\\
m\in \e(u)+\mathbb{Z},\\
n\in \e(v)+\mathbb{Z}.
\end{array}
\end{equation}
The Weyl vector
$\widehat{\rho}$ is defined by the set of  $r+1$
equations:
\begin{equation}    \label{rho def}
  2(\widehat{\rho}|\widehat{\alpha_i})=
(\widehat{\alpha_i}|\widehat{\alpha_i}),
\qquad \widehat{\alpha_i} - \text{simple roots of}\ \twg.
\end{equation}
The Weyl vector $\widehat{\rho}$ is not any more equal to
$(\rho,h^\lor,0)$
in the twisted case.
\begin{conjecture*}
Let $\twg$ be a twisted loop algebra with $\e(\fh)=0$.
Then the Weyl
vector ${\widehat \rho}$ defined by (\ref{rho def}) is given
by ${\widehat \rho}=({\widetilde \rho}, h^\lor,0)$, where
\begin{equation}             \label{rho twisted}
{\widetilde \rho}=\frac{1}{2} \sum_{\alpha \in \Delta_+}
(-1)^{p_\alpha} \alpha (1-2\e_\alpha)
\end{equation}
is a ``twisted rho'', $\e_\alpha \equiv \e(u_\alpha)$.
\end{conjecture*}
We will prove this conjecture in a special case only in
Section \ref{Determinant formula}.

A highest weight vector $\ket{\widehat \lambda}$
of weight $\widehat\lambda=(\lambda,k,0)$
is annihilated by $\widetilde \fn^+$ and it is
an eigenvector of the generators of the Cartan subalgebra
$\widetilde \fh$:
\begin{equation}
  \begin{aligned}
      {\widetilde \fn}^+ \ket{\widehat \lambda}&=0, \\
 D \ket{\widehat \lambda}&=0 ,
 \end{aligned}
 \qquad
 \begin{aligned}
 h_0 \ket{\widehat \lambda}&=\lambda(h) \ket{\widehat\lambda} ,
 \ h \in  \fh ,\\
 K \ket{\widehat \lambda}&=k \ket{\widehat\lambda} .
  \end{aligned}
\end{equation}

We would like to calculate the eigenvalue of $L^\fg_0$
on the highest weight vector $\ket{\widehat \lambda}$.
We do it in the case $\e(\fh)=0$.
In this case $\epsilon(u_\alpha)+\epsilon(u_{-\alpha})\in \mathbb{Z}$.
We will denote $\epsilon_\alpha=\epsilon(u_\alpha)$
(then $\epsilon(u^\alpha)=\epsilon_{-\alpha}$)
and choose
\begin{equation}
0 \le \epsilon_\alpha <1, \ \text{for} \ \alpha\in\Delta_+
\quad
\text{and}
\quad
\epsilon_{-\alpha}=-\epsilon_\alpha.
\end{equation}

In the Cartan--Weyl basis
the energy--momentum field is written as
\begin{equation}
   L^\fg =
\frac{1}{2(k+h^\lor)} \left(
\sum_{i=1}^r
\NO{h^i}{h_i}+
\sum_{\alpha \in \Delta}
\NO{u^\alpha}{u_\alpha}
\right).
\end{equation}
Using the formula (\ref{NOepsilon}) one can express
the energy--momentum zero mode as
\begin{equation}
\begin{aligned}
   L^\fg_0&=\frac{1}{2(k+h^\lor)} \left(
\sum_{i=1}^r \Bigg(
\sum_{n \in-1-\nz} h^i_n h_{i,-n}+
\sum_{n \in\nz}  h_{i,-n} h^i_n
\Bigg) \right.\\
&+
2\sum_{\alpha \in \Delta_+}
\Bigg(\sum_{m\in -\e_\alpha-\nz}
u^\alpha_m u_{\alpha,-m}
+(-1)^{p_\alpha}
\sum_{m\in 1-\e_\alpha+\nz}
u_{\alpha,-m} u^\alpha_m
\Bigg)\\
&+\left.
\sum_{\alpha \in \Delta_+}
(-1)^{p_\alpha} \Big(
k\,\e_\alpha(1-\e_\alpha)+
(1-2\,\e_\alpha)
{\com{u_\alpha}{u^\alpha}}_0
\Big)
\right).
\end{aligned}
\end{equation}
The $L^\fg_0$ eigenvalue is
\begin{equation}                        \label{Lg_0 eigenvalue}
  L^\fg_0 \ket{\widehat \lambda}=
\frac{1}{2(k+h^\lor)}\Big(
(\lambda | \lambda+2{\widetilde \rho})+
k \sum_{\alpha \in \Delta_+}
(-1)^{p_\alpha} \e_\alpha(1-\e_\alpha)
\Big) \ket{\widehat\lambda}\, ,
\end{equation}
where $\e_\alpha,\, \alpha \in \Delta_+$ are assumed to be in the range
$0 \le \e_\alpha <1$ and ${\widetilde \rho}$ is a twisted ``rho''
defined in (\ref{rho twisted}).

Next we would like to generalize the determinant formula to the
twisted case.
In the untwisted case the determinant formula
for the contravariant form on the weight space
with weight $\widehat\lambda-\widehat\eta$ of a Verma module
$R_{\widehat\lambda}$
with highest weight $\widehat\lambda$ is given by
(see \cite{Kac det, KW1})
\begin{equation}    \label{affine det formula}
{\det}_{\widehat \eta} (\widehat \lambda) =
\prod_{\widehat \alpha \in \widehat \Delta_+}
 \, \prod_{n \in \mathbb{N}} \,\,\left((\widehat \lambda + \widehat \rho
 |\widehat \alpha)-\frac{n}{2}(\widehat \alpha |
\widehat \alpha)\right)^
{q(\widehat \alpha,n)
P(\widehat \eta -n \widehat \alpha)
 \dim  \ag_{\widehat \alpha}},
\end{equation}
where $P(\tau)$ is the number of partitions of $\tau$
to the sum of positive roots, $\dim  \ag_{\widehat \alpha}$
is the dimension of the root space $\ag_{\widehat \alpha}$
associated to the root $\widehat \alpha$, and
$q(\widehat \alpha,n)=(-1)^{p(\widehat \alpha)(n+1)}$.

The above formula (\ref{affine det formula})
is valid also in the twisted
case, one has just to use the twisted set of positive roots and
the twisted $\widehat \rho$, defined by (\ref{rho def}).

When $\fg$ is a Lie superalgebra there are odd roots which lead
to a cancellation of some factors. If $\gamma$ is an odd
isotropic ($(\gamma|\gamma)=0$) root then the correspondent factor
does not depend on $n$, and one can evaluate the product on $n$
explicitly. If $\beta$ is odd, but not isotropic it is a half
of an even root, and then some factors corresponding to $\beta$
and to $2\beta$ cancel each other. As a result one expresses
the determinant formula
(\ref{affine det formula})
in the more explicit way (see \cite{KW1}):
\begin{equation}              \label{explicit affine det formula}
 \begin{aligned}
   {\det}_{\widehat{\eta}} (\widehat{\lambda}) &=
     (k+h^\lor)^{\sum_{m,n \in \mathbb{N}} P(\widehat{\eta}-(0,0,m n))}
     \prod_{n \in \mathbb{N}} \prod_{\widehat{\alpha}}
     \left((\widehat{\lambda}+\widehat{\rho}|\widehat{\alpha})-\tfrac{n}{2}
     (\widehat{\alpha}|\widehat{\alpha})
     \right)^{P(\widehat{\eta}-n\widehat{\alpha})} \times\\
   &  \times \prod_{n\in 1+2\nz} \prod_{\widehat{\beta}}
     \left(( \widehat{\lambda}+\widehat{\rho} |\widehat{\beta})-
     \tfrac{n}{2}(\widehat{\beta}|\widehat{\beta})\right)^
{P(\widehat{\eta}-
n\widehat{\beta})}
     \prod_{\widehat{\gamma}}  (\widehat{\lambda}+ \widehat{\rho}|
     \widehat{\gamma})^{P_{\widehat{\gamma}}
     (\widehat{\eta}-\widehat{\gamma})}\, ,
 \end{aligned}
 \end{equation}
where $\widehat{\alpha}=(\alpha,0,m)$
runs on even positive roots,
such that $\alpha \ne 0$
and $\half \widehat{\alpha}$ is not an odd root;
$\widehat{\beta}$ runs on odd positive roots, such that
$2\widehat{\beta}$ is an even root; $\widehat{\gamma}$
runs on odd positive roots, such that
$2\widehat{\gamma}$ is not a root
(then $(\widehat\gamma|\widehat\gamma)=0$);
$P_{\widehat{\gamma}}$ is a number of partitions
not involving $\widehat{\gamma}$.


\subsection{Superghost system}

\label{Superghost system}


Let $A$ be a finite dimensional vector superspace.
(In application to the quantum reduction
$A=\fg_>$ with flipped parity.)
Let $A_{\text{ch}}= A \oplus A^*$, define an even
skew--supersymmetric non--degenerate bilinear form
${<}{\,.\,}{,}{\,.\,}{>}_{\text{ch}}$ on $A_{\text{ch}}$ by
\begin{equation}
\begin{aligned}
{<}A, A{>}_{\text{ch}}&=0={<}A^*, A^*{>}_{\text{ch}} \\
{<}a, b^*{>}_{\text{ch}}&=-(-1)^{p(a)p(b^*)}{<}b^*,a{>}_{\text{ch}}=
b^*(a)
\end{aligned}
\end{equation}
for $a \in A$, $b^* \in A^*$.
We introduce a system of local fields
$\{ \ghg(z), \ghb(z) \}$
($\ghg \in A,
\ghb \in A^*$), called a superghost\footnote{
``Charged free superfermions'' in notation of \cite{KRW} and
\cite{KW1}; $b$-$c$ or $\beta$-$\gamma$ system in the physical
literature.
}
 system, subject to
the following \ope :
\begin{equation}
\ghg(z) \, \ghb(w)= \frac{1}
{z-w}\, {<}\ghg, \ghb{>}_{\text{ch}}.
\end{equation}
The vertex algebra of superghost fields is denoted by
$F(A_\text{ch})$.

Let $\{\ghg_i\}$ and $\{\ghb^i\}$ be the bases of $A$ and $A^*$
such that ${<}\ghg_i, \ghb^j{>}_{\text{ch}}=\delta_i^j$.
Then the superghost system decouples to a set of
mutually commuting ghost pairs:
\begin{equation}                         \label{gamma_beta_ope}
\ghg_i(z) \, \ghb^j(w)= \frac{1}
{z-w}\, \delta_i^j.
\end{equation}

A family of energy-momentum fields parameterized by
$\{\Delta(\ghb^j)\}$ is defined by
\begin{equation}                     \label{L gh}
  L^{\text{ch}}=-\sum_i \Delta(\ghb^i) \NO{\ghb^i}{\dd \ghg_i}+
\sum_i (1-\Delta(\ghb^i))\NO{\dd \ghb^i}{\ghg_i}.
\end{equation}
The field $L^{\text{ch}}(z)$ generates the Virasoro algebra
with central charge
\begin{equation}   \label{c_ch}
  c_{\text{ch}}=2 \sum_i (-1)^{p(\ghb_i)}
\left(6\Delta(\ghb^i)^2-6\Delta(\ghb^i)+1 \right).
\end{equation}
With respect to $L^{\text{ch}}$ the ghost field
$\ghb^i(z)$ (respectively $\ghg_i(z)$) is primary of conformal
dimension $\Delta(\ghb^i)$ (respectively $1- \Delta(\ghb^i)$).

The superghost system is called $\epsilon$--twisted
if its fields have the following mode expansions:
\begin{equation}
\ghg_i(z)=\sum_{n \in \epsilon(\ghg_i)+\Delta(\ghb^i)+\mathbb{Z}}
\ghg_{i,n} z^{-n-1+\Delta(\ghb^i)},
\qquad
\ghb^i(z)=\sum_{n \in \epsilon(\ghb^i)-\Delta(\ghb^i)+\mathbb{Z}}
\ghb^i_n z^{-n-\Delta(\ghb^i)}.
\end{equation}
The \ope\ (\ref{gamma_beta_ope}) can be written in terms of
commutation relations for the modes:
\begin{equation}
\com{\ghg_{i,m}}{\ghb^j_n}= \delta_i^j \delta_{n+m,0}\, ,
\quad
\begin{aligned}
m& \in \epsilon(\ghg_i)+\Delta(\ghb^i)+\mathbb{Z},\\
n & \in \epsilon(\ghb^j)-\Delta(\ghb^j)+\mathbb{Z} .
\end{aligned}
\end{equation}
We see from here, that $\e(\ghb^i)+\e(\ghg_i)\in \mathbb{Z}$.
We will 
choose $\e(\ghb^i)=-\e(\ghg_i)$.

A vacuum vector $\ket{0}_{\text{ch}}$ is defined by the set of conditions:
\begin{equation}    \label{0ch}
  \begin{aligned}
    \ghg_{i,m} \ket{0}_{\text{ch}} &=0, \quad m \ge \Delta(\ghb^i),\\
    \ghb^i_n \ket{0}_{\text{ch}} &=0, \quad n > -\Delta(\ghb^i).\\
  \end{aligned}
\end{equation}

The energy-momentum zero mode becomes
\begin{equation}
\begin{aligned}
L^{\text{ch}}_0&=\sum_i \left(
-(-1)^{p(\ghb_i)}
\frac{\e(\ghg_i)}{2}
(2\Delta(\ghb^i)+\e(\ghg_i)-1)-
\sum_{m \in -\Delta(\ghb^i)-\e(\ghg_i)-\nz} m\, \ghb^i_m
\ghg_{i,-m}\right. \\
&\left.-(-1)^{p(\ghb_i)}
\sum_{m \in -\Delta(\ghb^i)+1-\e(\ghg_i)+\nz} m \,
\ghg_{i,-m}\ghb^i_m
\right).
\end{aligned}
\end{equation}
The first term only contributes to the vacuum energy,
assuming $\epsilon$ is taken in the range $0\le \e(\ghg_i) < 1$
($\e(\ghg_i)=0$ corresponds to the untwisted case), i.e.~
\begin{equation}    \label{Lch_0 eigenvalue}
L^{\text{ch}}_0\ket{0}_{\text{ch}}=
\sum_i \left(
-(-1)^{p(\ghb_i)}
\frac{\e(\ghg_i)}{2}
(2\Delta(\ghb^i)+\e(\ghg_i)-1)
\right)
\ket{0}_{\text{ch}}\, .
\end{equation}

See also ref.~\cite{Eholzer:1997se} where the fermionic and bosonic ghost
systems are also discussed in the case of twisted boundary conditions.


\subsection{Neutral free superfermion system}

\label{Neutral free superfermion system}


Let $A=A_{\bar{0}} \oplus A_{\bar{1}}$ be a finite dimensional
superspace with a nondegenerate skew-symmetric even bilinear
form ${<}{\,.\,}{,}{\,.\,}{>}_{\text{ne}}$,
i.e.~it is skew-symmetric
on $A_{\bar{0}}$ and symmetric on $A_{\bar{1}}$
and
${<}A_{\bar{0}}, A_{\bar{1}}{>}_{\text{ne}}=0$.
A set of fields $\{ \psi(z) \}_{\psi \in A}$ is
called a system of neutral free
superfermions, if the fields satisfy the following \ope s:
\begin{equation}                    \label{ope_psi}
  \psi(z)\, \phi(w) \sim \frac{1}{z-w}
{<}\psi, \phi{>}_{\text{ne}}, \quad \psi, \phi \in A.
\end{equation}
The vertex algebra of neutral free superfermions is denoted by
$F(A_\text{ne})$.
In application to the quantum reduction
$A=\fg_{1/2}$ and the bilinear form
is defined by
\begin{equation}   \label{neutral bilinear form}
{<}u, v{>}_{\text{ne}}=
(f|\com{u}{v}),
\end{equation}
 where $u,v \in \fg_{1/2}$
and $f\in\fg_{-1}$ is a good element.

The energy--momentum field for the neutral free
superfermion system is
\begin{equation}    \label{L ne}
 L^{\text{ne}}=\frac{1}{2} \sum_i \NO{\dd \psi^i}{\psi_i}=
\frac{1}{2} \sum_i (-1)^{p(\psi_i)}
\NO{\psi_i}{\dd \psi^i},
\end{equation}
where $\{\psi_i\}$ and $\{\psi^i\}$ are dual bases of $A$:
\begin{equation}
{<}\psi_i , \psi^j{>}_{\text{ne}}=\delta_i^j \quad
\left(\text{then }
{<}\psi^i , \psi_j{>}_{\text{ne}}=-\delta_i^j (-1)^{p(\psi_i)}\right).
\end{equation}
The central charge of the Virasoro algebra generated by
$L^{\text{ne}}$ is
\begin{equation}
c_{\text{ne}}=-\frac{1}{2} \, \text{sdim} A.
\end{equation}
The neutral free superfermions are primary fields of conformal dimension
$1/2$ with respect to $L^{\text{ne}}$.

The superfermion 
fields have the following mode expansions:
\begin{equation}
\psi(z)=\sum_{n\in \epsilon(\psi)-1/2+\mathbb{Z}}
\psi_n z^{-n-1/2}.
\end{equation}
%
Commutation relations derived from (\ref{ope_psi})
read
\begin{equation}
\begin{split}
  \com{\psi_n}{\phi_m} = {<}\psi,\phi{>} \,\delta_{n+m,0} , \qquad
\psi , \phi \in A, \\
n\in \epsilon(\psi)-1/2+\mathbb{Z},\
m\in \epsilon(\phi)-1/2+\mathbb{Z}.
\end{split}
\end{equation}
The consistency condition on twistings is
\begin{equation}
  \e(\psi)+\e(\phi) \in \mathbb{Z},\ \text{if}\
 {<}\psi,\phi{>}\ne 0.
\end{equation}

The vacuum vector $\ket{0}_{\text{ne}}$ is defined
by the following conditions\footnote{
A different set of annihilation operators is chosen in \cite{KW2}.
}:
\begin{equation} \label{0ne}
  \psi_n \ket{0}_{\text{ne}}=0, \quad n>0,\ \psi \in A.
\end{equation}
If there are zero modes (it happens when
$\e(\psi)=1/2$ for some $\psi \in A$.)
one  has to specify their action on the vacuum vector
in order to complete the definition.

Next we would like to calculate the $L^{\text{ne}}_0$ eigenvalue
on the vacuum vector.
Using formula (\ref{NOepsilon}) one gets the following expression for the
energy--momentum zero mode:
\begin{equation}   \label{L_ne_0}
\begin{aligned}
  L^{\text{ne}}_0&=\frac{1}{2}\sum_i \left(
(-1)^{p(\psi_i)}  \!\!\! \!\!\!
\sum_{n\in -1/2+\e(\psi_i)-\nz}  \!\!\! \!\!\!
(n-\frac{1}{2})
\psi_{i,n} \psi^i_{-n} \right.\\
&\left. +  \!\!\!  \!\!\!
 \sum_{n \in 1/2+\e(\psi_i)+\nz}  \!\!\!  \!\!\!
(n-\frac{1}{2})\psi^i_{-n}\psi_{i,n}
-(-1)^{p(\psi_i)}\frac{1}{2}\e(\psi_i)\big(\e(\psi_i)-1\big)
 \right).
\end{aligned}
\end{equation}
If $-1/2 < \e(\psi_i) <1/2$ then  $\psi_i$
contributes  \mbox{
$-1/4(-1)^{p(\psi_i)}\e(\psi_i)\big(\e(\psi_i)-1\big)$
}
to the $L^{\text{ne}}_0$ vacuum eigenvalue.
The case $\e(\psi_i)=1/2$ should be treated separately.
Since $\sum_i \psi_{i,0} \psi^i_0
=-\sum_i(-1)^{p(\psi_i)}\psi^i_0 \psi_{i,0}$
the first term in (\ref{L_ne_0}) contributes
$-1/8(-1)^{p(\psi_i)}$ to the eigenvalue, and the overall
contribution is $-1/16(-1)^{p(\psi_i)}$. Finally we have
\begin{equation}
  L^{\text{ne}}_0 \ket{0}_{\text{ne}}=
\left(\sum_i h^{\text{ne}}_i \right) \ket{0}_{\text{ne}},
\end{equation}
where
\begin{equation}    \label{Lne_0 eigenvalue}
h^{\text{ne}}_i=\left\{
\begin{aligned}
&-1/4(-1)^{p(\psi_i)}\e(\psi_i)\big(\e(\psi_i)-1\big),
\ &-1/2 < \e(\psi_i) < 1/2,\\
&-1/16(-1)^{p(\psi_i)},\ &\e(\psi_i) = 1/2.
\end{aligned}
\right.
\end{equation}
In particular the $L^{\text{ne}}_0$ vacuum eigenvalue
is equal to zero when all the superfermions are untwisted,
and equal to $-1/16 \,\text{sdim}A$ when $\e(\psi)=1/2$
for all $\psi \in A$.


\section{Quantum reduction}

\label{Quantum reduction}

\setcounter{equation}{0}


The details of the construction can be found in
\cite{KRW,KW1}, we reproduce here only main points and results.


\subsection{Homology complex}

\label{BRST cohomology}


Let $\fg$ be a simple Lie superalgebra with a good gradation on it,
generated by an element $x\in\fg$,
as described in Section \ref{Gradation on a Lie superalgebra}.
Then one introduces three types of vertex algebras:
the affine vertex algebra $V_k(\fg)$
(Section~\ref{Affine vertex algebra}),
the superghost algebra $F(A_{\text{ch}})$
(Section~\ref{Superghost system}),
and the superfermion algebra $F(A_{\text{ne}})$
(Section~\ref{Neutral free superfermion system}).

If the Cartan subalgebra $\fh$ is untwisted
(this is always assumed in the current paper),
then the root elements have a well defined twisting.
If in addition one chooses $x\in\fh$, then the root elements
have also a well defined grading by $x$, and it is convenient
to use the Cartan--Weyl basis of $\fg$ in the calculations.

Let $\Delta_+$ be the set of positive roots compatible
with the chosen gradation, i.e. $\alpha(x)\ge 0$ if $\alpha\in
\Delta_+$.
Let $\Delta_{j}$ (respectively $\Delta_>$) be a set of roots corresponding to
$\fg_{j}$ (respectively $\fg_>$) as defined in (\ref{Delta_j}) and in (\ref{Delta greater}).

The base space for the superghost
algebra is the $\fg_>$ space with flipped parity.
One introduces a $\ghb$-$\ghg$ pair for each $\alpha\in \Delta_>$.
The parity of the $\ghb$-$\ghg$ pair is $p(\ghb^\alpha)=p(u_\alpha)+1$,
i.e.~it is odd, if $u_\alpha$ is even, and even if $u_\alpha$ is odd.
The parameter $\Delta(\ghb^\alpha)=
1-\Delta(\ghg_\alpha)$
is chosen to be equal to the
gradation:
\begin{equation}
\Delta(\ghb^\alpha)=j
\qquad
\text{if } u_\alpha \in \fg_j.
\end{equation}
Then the central charge (\ref{c_ch}) of the superghost Virasoro
algebra becomes
\begin{equation}
  c_{\text{ch}}=-2 \sum_{j \in \frac{1}{2}\mathbb{N}} \text{sdim}\,\fg_{j}
\left(6j^2-6j+1 \right).
\end{equation}

For any basis element $u_\alpha,\,{\alpha\in \Delta_{1/2}}$
of $\fg_{1/2}$ one
should also add a neutral superfermion
with the same parity as $u_\alpha$.
The bilinear form on $\fg_{1/2}$ is given
by~(\ref{neutral bilinear form}).

Now we are ready to introduce an odd field
$d(z)$ in the vertex algebra
${\cal{C}}(\fg,x,k)=V_k(\fg)\otimes F(A_{\text{ch}})
\otimes F(A_{\text{ne}})$:
\begin{equation}        \label{d(z)}
  d=\sum_{\alpha \in \Delta_>}
(-1)^{p_\alpha}\,  u_\alpha \ghb^\alpha
- \half \sum_{\alpha, \beta, \gamma \in \Delta_>} \!\!\!\!
(-1)^{p_{\alpha} p_{\gamma}}
\str_{{\alpha} {\beta}}^{\, \gamma \, }\,
\ghg_{\gamma} \ghb^{\alpha} \ghb^{\beta}
+\sum_{\alpha \in \Delta_>} (f|u_\alpha) \ghb^\alpha +
\sum_{\alpha \in \Delta_{1/2}} \ghb^\alpha \psi_\alpha,
\end{equation}
where $p_\alpha=p(u_\alpha)$ and $\str_{{\alpha} {\beta}}^{\, \gamma \, }$
are structure constants
of $\fg$:
\begin{equation}
  \com{u_\alpha}{u_\beta}= \str_{\alpha \beta}^{\, \gamma \, } u_\gamma.
\end{equation}
The normal ordering is not necessary since all the fields are
commutative in the $d(z)$ monomials.

The key feature of the field $d(z)$ is that the singular part
of its \ope\ with itself vanishes:
\begin{equation}
  \label{eq:dd ope}
  d(z) d(w)= \text{regular in }(z-w)\, .
\end{equation}
The proof can be found in \cite{KRW} (Theorem 2.1).

Define an operator $d_0$ on ${\cal{C}}(\fg,x,k)$ to be
the first order pole in the \ope\ of $d(z)$ with a field
from ${\cal{C}}$:
\begin{equation}
  \label{eq:define d0}
  d(z)\phi(w)=\ldots + \frac{(d_0 \phi)(w)}{z-w}+\ldots 
\end{equation}
One can deduce from the associativity condition of
\ope s that $d_0$ is an odd derivation of an \ope , i.e.
\begin{equation}
  \label{eq:d0 derivation}
  d_0 [ \phi_1 \phi_2 ]^{(q)} = [(d_0 \phi_1) \phi_2 ]^{(q)}+
(-1)^{p(\phi_1)} [\phi_1 (d_0 \phi_2) ]^{(q)} \, ,
\end{equation}
where $[ A B ]^{(q)}$ is a pole of order $q$ in the
\ope\ of $A$ with $B$:
\begin{equation}
  A(z) B(w) = \sum_{l=-N(A,B)+\nz}
[ A B ]^{(-l)} (z-w)^l .
\end{equation}
In particular $d_0$ is an odd derivation with respect to
the normal ordered product $\NO{\phi_1}{\phi_2}=[ \phi_1 \phi_2 ]^{(0)}$.

The following crucial feature of $d_0$:
\begin{equation}
  d_0^2=0
\end{equation}
is an immediate consequence of (\ref{eq:dd ope}) and (\ref{eq:d0 derivation}).

Next one builds a homology complex $({\cal{C}}(\fg,x,k),d_0)$
(``BRST cohomology'' in physical literature)
of vertex algebra ${\cal{C}}$ with respect to
$d_0$. The homology of the complex
\begin{equation}
  H( {\cal{C}}, d_0)= \text{Ker}\, d_0 / \text{Im}\, d_0
\end{equation}
is a vertex algebra, the quantum reduction
of $\fg$ with respect to $x$. It is denoted $W_k(\fg,x)$.

A charge can be assigned to the fields in ${\cal{C}}$:
\begin{equation}   \label{charge}
  \text{charge}\,V_k(\fg)=0, \quad
  \text{charge}\,F(A_{\text{ne}})=0, \quad
  \text{charge}\,\ghb =-1, \quad
\text{charge}\,\ghg =1.
\end{equation}
Then the vertex algebra ${\cal{C}}(\fg,x,k)$ has charge decomposition
\begin{equation}     \label{charge decomposition}
  {\cal{C}}(\fg,x,k)=\bigoplus_{m\in\mathbb{Z}} {\cal{C}}_m\, .
\end{equation}
The field $d(z)$ has charge $-1$, hence $d_0$ lowers the charge by 1:
$d_0({\cal{C}}_m)\subset {\cal{C}}_{m-1}$ and
$({\cal{C}}(\fg,x,k), d_0)$ is a $\mathbb{Z}$-graded homology complex.


\subsection{Twist gluing}

\label{Twist gluing}


We have three commuting vertex algebras: $V_k(\fg)$,
$F(A_{\text{ch}})$ and $F(A_{\text{ne}})$. Each of them
can be twisted in a self-consistent way as described in
Section~\ref{Ingredients}. However in the quantum reduction procedure
these twistings should be related. The restrictions come from
the demand that the field $d(z)$ should be untwisted.
Denote $\e_\alpha=\e(u_\alpha)$, where $\alpha$ is a positive root of $\fg$.
Then (since we consider the case, when the Cartan subalgebra is
untwisted) we choose $\e(u_{-\alpha})=-\e_\alpha$ for
$\alpha \in \Delta_+$. From the first term in $d(z)$ (\ref{d(z)})
we see that $\e(\ghb^\alpha)=-\e_\alpha$ and therefore
(see section~\ref{Superghost system}) $\e(\ghg_\alpha)=\e_\alpha$.
Let $e\in\fg$ be an element dual to $f$, then we obtain from the third
term in (\ref{d(z)}),
that the ghost field $\ghb$ associated to $e$ is untwisted and therefore
$e$ and $f$ themselves are also untwisted.
From the last term one gets that $\e(\psi_\alpha)=\e_\alpha$.
Finally, we conclude that all the possible twistings are parameterized by
$r=\text{rank}\, \fg$ numbers $\e_\alpha$, $\alpha$ are the simple roots of $\fg$,
modulo the condition that $\e(e)=0$.

There are two cases of particular interest in physics:
the Neveu--Schwarz (NS) sector and the Ramond sector.
These sectors may be defined for a good gradation on any
Lie superalgebra $\fg$. The NS sector is simply the untwisted
case: $\e(u)=0,\ \forall\, u\in \fg$. The Ramond sector is
defined by the following twistings:
\begin{equation}
\e(u)=\left\{
\begin{array}{lll}
0, & u\in \fg_j, & j\in \mathbb{Z},\\
1/2,& u\in \fg_j, & j\in 1/2+\mathbb{Z}.
\end{array}
\right.
\end{equation}


\subsection{Structure of the W-algebra}

\label{Structure of the W-algebra}


Here we reproduce the results of \cite{KRW,KW1} on the structure
of $W_k(\fg,x)$. The first fact is that the Virasoro algebra
is always contained in $W_k(\fg,x)$. It is generated by the field $L(z)$:
\begin{equation}
  \label{eq:L}
  L=L^{\fg}+L^{\text{ch}}+L^{\text{ne}}+\dd x,
\end{equation}
where $L^{\fg},L^{\text{ch}},L^{\text{ne}}$ are the energy--momentum fields
from Sections~\ref{Affine vertex algebra}, \ref{Superghost system},
\ref{Neutral free superfermion system} respectively.
Due to the $\dd x$ term
the conformal dimensions of affine currents are shifted from 1
with respect to the Virasoro field $L(z)$:
the field $u(z)$ is of dimension $1-j$ if $u\in \fg_j$.
Then one can easily check that $d(z)$ is of dimension 1 with respect
to $L(z)$ and therefore $d_0 L =0$ and that
$L$ is not in $\text{Im}\,d_0$. The central charge of the Virasoro
algebra generated by $L(z)$ is
\begin{equation}
  \label{eq:central charge}
\begin{aligned}
  c&=c_\fg+c_{\text{ch}}+c_{\text{ne}}-12k(x|x)\\
   &=\frac{k\,\text{sdim}\fg}{k+h^\lor}\,
-2 \sum_{j>0} \text{sdim}\,\fg_{j}
\left(6j^2-6j+1 \right)
-\frac{1}{2} \, \text{sdim}\, \fg_{1/2}
-12k(x|x)
\end{aligned}
\end{equation}

The structure of the W-algebra $W_k(\fg,x)$ is described
in the main theorem of \cite{KW1} (Theorem 4.1).
Let $\fg^f=\{u\in \fg | \com{u}{f}=0\}$ be the centralizer
of $f$ in $\fg$. Denote
\begin{equation}
  \label{eq:ja}
  J^{(v)}=v+\sum_{\alpha , \beta \in \Delta_>}
(-1)^{p_{\beta}} \str_{v \alpha}^{\, \beta \,}
\NO{\ghg_{\beta}}{\ghb^{\alpha}},
\end{equation}
where $v\in\fg$ and $\str_{v \alpha}^{\, \beta \,}$ are the structure
constants of $\fg$: $\com{v}{u_{\alpha}}=\sum_{\beta \in \Delta}
\str_{v \alpha}^{\, \beta \,} u_{\beta}$.
The theorem states that
\begin{enumerate}

\item

The only nontrivial homology
lies in ${\cal{C}}_0$:
\begin{equation}
\begin{aligned}
  H_l({\cal{C}}(\fg,x,k),d_0)&=0, \ \text{if } l \ne 0,\\
  H_0({\cal{C}}(\fg,x,k),d_0)&=W_k(\fg,x);
\end{aligned}
\end{equation}

\item

The W-algebra $W_k(\fg,x)$ is strongly generated by
homology classes of fields
$J^{\{a_i\}}$ where $a_i \in \fg^f, i=1,2,\ldots,\text{dim}\,\fg^f$
is a basis of $\fg^f$ compatible with the gradation;

\item

If $a\in \fg_{-j}$ then
the field $J^{\{a\}}$ is of dimension $1+j$ with respect to $L(z)$
and $J^{\{a\}}$ is equal to $J^{(a)}$ plus a linear combination
of normal ordered products of the fields
$J^{(b)}$, where $b\in \fg_{-s}, \ 0 \le s <j$, the fields
$\psi_\alpha, \ \alpha \in \Delta_{1/2}$ and their derivatives.

\end{enumerate}

The Virasoro field $L(z)$~(\ref{eq:L}) is in the same homology class
as $J^{\{f\}}(z)$, and since $f\in \fg^f$ the field $L(z)$
is always part of the W-algebra $W_k(\fg,x)$.

In the case of a good gradation $\fg^f \subset \fg_{\le}$,
therefore the conformal dimensions of the generating fields are greater
or equal to 1. This is in agreement with the result
of~\cite{Goddard:1988wv}, that dimension $1/2$ fields can be
factored out from a W-algebra.


\subsection{Highest weight modules of the W-algebra}

\label{Representation theory}


In this section we are going to discuss highest weight
representations of W-algebras $W_k(\fg,x)$ in a framework of
quantum reduction. The discussion applies to a general
twisted case.

A highest weight vector of the vertex algebra
${\cal{C}}(\fg,x,k)=V_k(\fg)
\otimes F(A_{\text{ch}})
\otimes F(A_{\text{ne}})$ is given by
\begin{equation}
  \label{eq:hw state}
 \ket{\lambda}_k= \ket{\widehat \lambda} \times \ket{0}_{\text{ch}}
\times \ket{0}_{\text{ne}}.
\end{equation}
The full highest weight module $Q_k(\lambda)$ is obtained by
applying affine, superghost and superfermion
creation operators to the highest weight vector.

One introduces highest weight representations of $W_k(\fg,x)$
in the following way. First define the mode expansions
of the generating fields:
\begin{equation}               \label{W mode expansion}
  J^{\{u\}}(z)=\sum_{n \in -\Delta(u)+\e(u)+\mathbb{Z}}
 J^{\{u\}}_n z^{-n-\Delta(u)},
\end{equation}
where $\Delta(u)=1+j$, if $u\in\fg_{-j}$,
is the conformal dimension of the field
$J^{\{u\}}(z)$ with respect to the Virasoro
field~(\ref{eq:L}).
The W-algebra  highest weight vector is 
annihilated by positive modes of
all the fields forming the W-algebra
($W_n, n>0, W\in W_k(\fg,x)$).
One should also treat the zero modes.
First choose a set of mutually commuting (in strong or weak sense)
zero modes which is called a set of Cartan generators.
Two operators are called commutative in weak sense if their
commutator is zero modulo terms which annihilate highest weight
vectors. (An example of the W-algebra with Cartan generators
commutative in weak sense is studied in \cite{Noyvert:2002mc}.)
The highest weight vector is an eigenvector of the Cartan generators
and it is labelled by the eigenvalues of the Cartan operators.
Some of the non Cartan zero modes should also annihilate the
highest weight vector.

One can check that the positive modes of the fields
$J^{\{u\}}(z)$ annihilate the vector $\ket{\lambda}_k$
defined in (\ref{eq:hw state}):
\begin{equation}
  J^{\{u\}}_n \ket{\lambda}_k=0, \qquad n>0,\ u \in \fg^f.
\end{equation}
So this vector can be chosen as a highest weight vector
of the W-algebra.

It is easy to see that the highest weight vector is
$d_0$ closed:
\begin{equation}
  d_0  \ket{\lambda}_k=0 .
\end{equation}
To get the W-algebra module $M_k(\lambda)$ one should take the $d_0$
homology of the ${\cal{C}}$ module:
$M_k(\lambda)=H(Q_k(\lambda), d_0)$.

The charge decomposition (\ref{charge decomposition})
is extended to the $Q_k(\lambda)$ module by a field--state
correspondence (charge of the highest weight vector
$\ket{\lambda}_k$ is taken to be zero). Then again only the
zero charge homology of the complex $(Q_k(\lambda), d_0)$
is nontrivial (see Theorem~6.2 of \cite{KW1}), and one has
\begin{equation}
M_k(\lambda)=H_0(Q_k(\lambda), d_0).
\end{equation}

Suppose there is a singular vector $\ket{\widehat s}$ in a highest
weight module $R_{\widehat \lambda}$ of the twisted loop algebra
$\twg$. Then the vector
\begin{equation}
\ket{s}_k=\ket{\widehat s}\times \ket{0}_{\text{ch}}
\times \ket{0}_{\text{ne}}
\end{equation}
is a singular vector in the ${\mathcal C}(\fg,x,k)$ algebra module
$Q_k(\lambda)$. This vector is $d_0$ closed  and is annihilated by
positive modes of the $W_k(\fg,x)$ algebra generators, therefore
(if it is not $d_0$ exact) the vector $\ket{s}_k$ is also a
singular vector in the W-algebra module $M_k(\lambda)$.


\section{Minimal W-algebras}

\label{Minimal W-algebras}

\setcounter{equation}{0}


\subsection{Structure}


In the case of minimal gradation (see Section~\ref{Gradation on a
Lie superalgebra}) $\fg^f$ can be easily described:
\begin{equation}
  \fg^f=\mathbb{C}f \oplus \fg_{-1/2} \oplus \fg_{0}^\natural ,
\end{equation}
where $\fg_{0}^\natural=\{u\in \fg_0 | (u|x)=0\}$ is a subspace of
$\fg_0$, orthogonal to $x$ with respect to the even invariant
bilinear form: $\fg_0=\mathbb{C} x \oplus \fg_{0}^\natural$. Then
the $W_k(\fg,x)$ algebra is generated by a Virasoro field, a
number of dimension $3/2$ fields and a number of dimension 1
fields. In the case of minimal gradation $(x|x)=1/2$, so one can
rewrite the formula for the central charge (\ref{eq:central charge})
in the simple form:
\begin{equation}                              \label{c_minimal_gradation}
c=\frac{k\,\text{sdim}\fg}{k+h^\lor} -6k
+\frac{1}{2} \, \text{sdim}\, \fg_{1/2}-2.
\end{equation}
The dimension 1 fields are given by
(Theorem 2.1 of~\cite{KW1})
\begin{equation}  \label{Jdim1}
  J^{\{v\}}=J^{(v)}-\frac{1}{2}
\sum_{\alpha, \beta \in \Delta_{1/2}} (-1)^{p_\beta} \str_{v
\alpha}^{\, \beta \,} \NO{\psi_\beta}{\psi^\alpha}, \quad
v\in\fg_{0}^\natural \, ,
\end{equation}
and the dimension $3/2$ fields are given by
\begin{equation}    \label{Gdim32}
\begin{aligned}
  G^{\{v\}}&=J^{(v)}
-\frac{(-1)^{p_v}}{3}\sum_{\alpha,\beta \in \Delta_{1/2}}
{:}\psi^\alpha \psi^\beta \psi_
{\com{u_\beta}{\com{u_\alpha}{v}}}{:}+
\sum_{\alpha \in \Delta_{1/2}} J^{(\com{v}{u_\alpha})}{\psi^\alpha}\\
&-\sum_{\alpha \in \Delta_{1/2}} \big( k (v|u_\alpha)+
  \text{str}_{\fg_>}(\text{ad}v\,\text{ad}u_\alpha)
\big) \dd \psi^\alpha , \quad v \in \fg_{-1/2} \, ,
\end{aligned}
\end{equation}
where $\psi_u$ means $\sum_\alpha a_\alpha \psi_\alpha$ if
$u=\sum_\alpha a_\alpha u_\alpha$.

The explicit form of \ope s of the W-algebra, corresponding to the
minimal gradation, is given in Theorem 5.1 of~\cite{KW1}.
The dimension--1 fields form a subalgebra with \ope s:
\begin{equation}
J^{\{a\}}(z) J^{\{b\}}(w)=
\frac{(a|b)(k+\frac{1}{2} h^\lor)-
\frac{1}{4}\,\text{str}_{\fg_0
}(\text{ad}a\,\text{ad}b)}
{(z-w)^2}
+\frac{J^{\{\com{a}{b}\}}(w)}{z-w}\, ,
\end{equation}
where $k$ is the level of $\fg$ and $a, b \in \fg_0^\natural$.
If $\fg_0^\natural$ is simple then the subalgebra is an affine vertex algebra
in the definition of Section~\ref{Affine vertex algebra}.
The \ope\ of $J$ and $G$ is:
\begin{equation}
J^{\{v\}}(z)\, G^{\{u\}}(w)=\frac{G^{\{\com{v}{u}\}}}{z-w} \, ,
\end{equation}
here $v\in \fg_0^\natural$ and $u\in \fg_{-1/2}$.
The fusion rule for two dimension $3/2$ fields:
\begin{equation}
G\times G=L+ J+{:}J J{:},
\end{equation}
and their explicit \ope\ can be found in Theorem 5.1(e) of~\cite{KW1}.

The algebras of ``minimal'' type were studied from a
different point of view in \cite{Fradkin:1992bz,Fradkin:1992km}.
The classification of minimal gradations on simple Lie
superalgebras (see tables in Proposition 4.1 of~\cite{KRW}) gives
also a classification of minimal W-algebras.

We have the following set of operators generating the W-algebra:
\begin{equation}\label{LGJ}
    \begin{aligned}
    L_n,\quad & n\in \mathbb{Z},\\
    G_r^{\{u\}},\quad &u\in\fg_{-1/2},\, r\in 1/2+\e(u)+\mathbb{Z},\\
    J_m^{\{v\}}, \quad & v\in \fg_0^\natural ,\, m\in \e(v)+\mathbb{Z}.
\end{aligned}
\end{equation}
The Cartan subalgebra is a span of
\begin{equation}
\{L_0\} \cup \{J_0^{\{v\}}|
v\in \fh^\natural \},
\end{equation}
where $\fh^\natural$ is a subspace of
$\fh$, orthogonal to $x$ with respect to the bilinear form $(.|.)$:
\mbox{$\fh=\mathbb{C} x \oplus \fh^\natural$.}

The Cartan generators act diagonally on the other generators.
One can introduce roots of the W-algebra. They consist of two
components: the first is an eigenvalue of
$\text{ad}\,J^{\{v\}}_0,\ v\in \fh^\natural$,
the second is an eigenvalue of $\text{ad}\,L_0$.
Then the root system $\Delta_W$
of the minimal W-algebra $W_k(\fg,x)$ is
a disjoint union of
\begin{equation}
\begin{array}{c}
\{(\alpha,m) \,|\, \alpha \in \Delta_0, m \in \e_\alpha
+\mathbb{Z}\},\\
\{(\alpha^\natural,m) \,|\, \alpha \in \Delta_{1/2},
m \in \frac{1}{2}+\e_\alpha+\mathbb{Z}\},
\{(0,m) \, |\, m \in \mathbb{Z}\},
\end{array}
\end{equation}
where
the multiplicity of the last set is $r=\text{rank}\,\fg$,
and $\alpha^\natural$ is the orthogonal projection of $\alpha$:
$\alpha^\natural=\alpha-\frac{1}{2} (\alpha|\theta)$.

There exists an anti-involution $\omega$ on a minimal W-algebra.
It is defined as
\begin{equation}           \label{anti-involution}
\begin{aligned}
\omega(L_n)&=L_{-n},&&\\
\omega(J^{\{v\}}_n)&=J^{\{v\}}_{-n}, \quad &v &\in \fh^\natural,\\
\omega(J^{\{v_\alpha\}}_n)&=J^{\{v_{-\alpha}\}}_{-n}, \quad &\alpha &\in \Delta_0
,\\
\omega(G^{\{u_\beta\}}_n)&=G^{\{u_{-\theta-\beta}\}}_{-n}, \quad &\beta &\in \Delta_{-1/2}.
\end{aligned}
\end{equation}
The proof that it is indeed an anti-involution can be found in Section 6 of~\cite{KW1}.


\subsection{Representation theory}

\label{minimal representation theory}


A highest weight vector $\ket{\Lambda,h}$ is defined by
\begin{equation}
\begin{aligned}
L_n, G_n^{\{u\}}, J_n^{\{v\}} \ket{\Lambda,h} &=0, n>0,\\
L_0 \ket{\Lambda,h}&=h \ket{\Lambda,h}\\
J_0^{\{v\}} \ket{\Lambda,h}&= \Lambda(v) \ket{\Lambda,h}, v\in\fh^\natural .
\end{aligned}
\end{equation}
The zero modes which are not in the Cartan subalgebra (if there
are such modes) should be treated separately.
We want to split $\Delta_0$ and $\Delta_{1/2}$ to positive
and negative parts. The splitting of $\Delta_0$ is naturally given
by a set of positive roots of $\fg$:
$\Delta_0^+=
\Delta_+ \cap \Delta_0$. To split $\Delta_{1/2}$ one has
to choose $h_0 \in \fh^\natural$ such that
\begin{equation}                           \label{Delta splitting}
\begin{aligned}
\alpha(h_0)> 0,\quad & \forall \alpha \in \Delta_0^+ ,\\
\alpha(h_0) \ne 0, \quad & \forall \alpha \in \Delta_{1/2}\
(\text{except } \alpha = \theta/2).
\end{aligned}
\end{equation}
We introduce
\begin{equation}
\begin{aligned}
\Delta_j^+&=\{\alpha\,|\,\alpha \in \Delta_j \
\text{and}\ \alpha(h_0)>0\},\\
\Delta_{j}^-&=\{\alpha\,|\,\alpha \in \Delta_j \
\text{and}\ \alpha(h_0)<0\},
\end{aligned}
\qquad j=-1/2,0,1/2.
\end{equation}
The root $\frac{1}{2}\theta$ (if there is such a root)
does not belong to  $\Delta_{1/2}^+ \cup \Delta_{1/2}^-$.
Note that there can be a few choices of the $\Delta_{1/2}$
splittings corresponding to the same $\Delta_0^+$.

Now we complete the definition of a highest weight vector:
\begin{equation}      \label{J0G0}
\begin{aligned}
J_0^{\{\alpha\}} \ket{\Lambda,h} &=0, \quad \alpha \in \Delta_0^+,
\e_{\alpha}=0, \\
G_0^{\{\beta\}} \ket{\Lambda,h} &=0, \quad \beta \in \Delta^+_{-1/2},
\e_\beta=1/2,
\end{aligned}
\end{equation}
where we denote $J^{\{\alpha\}}=J^{\{v_\alpha\}},
G^{\{\beta\}}=G^{\{u_\beta\}}$.
If $\theta/2 \in \Delta$ and $\e_{\theta/2}=1/2$,
then there is a fermionic operator  $G^{\{{-\theta/2}\}}_0$.
It commutes with Cartan generators and therefore
the highest weight vector $\ket{\Lambda,h}$ is its eigenvector:
\begin{equation}                   \label{G0 hw}
G^{\{{-\theta/2}\}}_0 \ket{\Lambda,h} = g(\Lambda,h) \ket{\Lambda,h}.
\end{equation}
The eigenvalue $g(\Lambda,h)$ can be calculated from the
$\com{G^{\{{-\theta/2}\}}_0}{G^{\{{-\theta/2}\}}_0}$
bracket.

In the quantum reduction procedure the highest weight vector
$\ket{\Lambda,h}$ is given by
\begin{equation}
\ket{\Lambda,h}=\ket{\widehat \lambda} \times \ket{0}_\text{ch} \times
\ket{0}_\text{ne},
\end{equation}
where $\Lambda$ and $h$ are functions of $\lambda$ and $k$, which are
calculated below. But first we should complete the definition
of the superfermion vacuum $\ket{0}_\text{ne}$ (see~\ref{0ne})
by specifying the action of superfermion zero modes on it
in agreement with the second line in~(\ref{J0G0}).
From the last term in~(\ref{Gdim32}) and since
the dual superfermion is $\psi^\alpha=\psi_{\theta-\alpha}$
we get that
\begin{equation}
\psi_{\alpha,0} \ket{0}_\text{ne}=0, \quad \alpha \in
\Delta^+_{1/2}, \, \e_\alpha=1/2.
\end{equation}
It is easy to see that this is also sufficient condition for
the second equation in~(\ref{J0G0}).
The fermion $\psi_{\theta/2}$ is self dual, so
$\com{\psi^{\theta/2}_0}{\psi^{\theta/2}_0}=1$
(if $u_{\theta/2}$ is appropriately normalized),
then $\psi^{\theta/2}_0 \ket{0}_\text{ne}=
\frac{1}{\sqrt{2}} \ket{0}_\text{ne}$.

Now we are ready to calculate the $\Lambda(\lambda,k)$ and $h(\lambda,k)$
dependence. From~(\ref{eq:L}) we get
\begin{equation}
h=h^\fg+h^{\text{ch}}+h^{\text{ne}}-\frac{1}{2}(\lambda|\theta),
\end{equation}
where (see (\ref{Lg_0 eigenvalue}), (\ref{Lch_0 eigenvalue})
and (\ref{Lne_0 eigenvalue})) 
\begin{align}
h^\fg&=\frac{1}{2(k+h^\lor)}\Big(
(\lambda | \lambda+2\widetilde\rho)+
\sum_{\alpha \in \Delta_+}
(-1)^{p_\alpha} k\,\e_\alpha(1-\e_\alpha)
 \Big),
 \\
 \label{h_ch}
h^\text{ch}&=\frac{1}{2}\sum_{\alpha \in \Delta_{1/2}}
(-1)^{p_\alpha} \e_\alpha^2,
\\
\label{h ne}
h^\text{ne} &= -\frac{1}{4} \sum_{\alpha \in \Delta_{1/2}}
(-1)^{p_\alpha}(\e_\alpha-1)(\e_\alpha-2 \Theta(\e_\alpha-1/2)).
\end{align}
All $\e_\alpha$ are assumed to be in the range $0 \le \e_\alpha
<1$; $\Theta(x)$ is the step function:
\begin{equation}
\Theta(x)=\left\{
\begin{array}{ll}
1,&x>0\\
1/2,&x=0\\
0,&x<0
\end{array}
\right.
\end{equation}
To calculate $\Lambda(v)$ we rewrite (\ref{Jdim1}) for the case $v\in
\fh^\natural$:
\begin{equation}
J^{\{v\}}=v+\sum_{\alpha\in\Delta_>} (-1)^{p_\alpha} \alpha(v)
{:}\ghg_\alpha \ghb^\alpha{:}
-\frac{1}{2}\sum_{\alpha\in\Delta_{1/2}} (-1)^{p_\alpha} \alpha(v)
{:}\psi_\alpha \psi^\alpha{:}, \quad v\in\fh^\natural.
\end{equation}
Then using the formula~(\ref{NOepsilon}) and the definitions of
$\ket{0}_\text{ch}$ (\ref{0ch}) and $\ket{0}_\text{ne}$ (\ref{0ne})
\begin{equation}    \label{Lambda lambda}
\Lambda(v)=\lambda^\natural(v)-\frac{1}{2}\sum_{\alpha\in\Delta_{1/2}}
(-1)^{p_\alpha} \alpha(v) (\e_\alpha+\varkappa_\alpha),
\end{equation}
where again $0\le\e_\alpha<1$, and $\varkappa_\alpha$ is the
eigenvalue of the
$\psi_{\alpha,-1/2+\e_\alpha}\psi^\alpha_{1/2-\e_\alpha}$
operator:
\begin{equation}           \label{varkappa}
\varkappa_\alpha=\left\{
\begin{array}{ll}
0,&0\le\e_\alpha<1/2, \\
0,&\e_\alpha=1/2,\ \alpha\in\Delta_{1/2}^-,\\
1,&\e_\alpha=1/2,\ \alpha\in\Delta_{1/2}^+,\\
1,&1/2<\e_\alpha<1.
\end{array}
\right.
\end{equation}
and $\lambda^\natural$ is the projection of
$\lambda$ orthogonal to $\theta$:
\begin{equation}
\lambda=\lambda^\natural+\frac{1}{2}(\lambda|\theta) \theta,\quad
(\lambda^\natural|\theta)=0.
\end{equation}
Then $\lambda(v)=\lambda^\natural(v)$, if $v\in\fh^\natural$.

We are interested mainly in two twistings:
\begin{align}
                       \label{NS sector minimal}
\text{NS sector:}\qquad&\e_\alpha=0 \ \forall \alpha \in \Delta,
\\[3pt]
                       \label{Ramond sector minimal}
\text{Ramond sector:}\qquad&\e_\alpha=\left\{
\begin{array}{ll}
1/2,& \alpha\in\Delta_{1/2},\\
0,& \alpha\in\Delta_0 \text{ or } \alpha=\theta.
\end{array}
\right.
\end{align}
In the NS sector the modes of dimension-$3/2$ generators are in $1/2+\mathbb{Z}$,
the modes of other generators are integer. In the Ramond sector all the modes are
integer.
In these cases the $\lambda, k$ dependence of weights $\Lambda$ and $h$
is easily expressed:
\begin{align}
\label{Lambda h NS}
\text{NS: }&
\begin{aligned}
\Lambda&=\lambda^\natural,\\
h&=\frac{(\lambda|\lambda+2\rho)}{2(k+h^\lor)}-\frac{1}{2}(\lambda|\theta),
\end{aligned}
\\[3pt]
\label{Lambda h Ramond}
\text{R: }&
\begin{aligned}
\Lambda&=\lambda^\natural-\rho_{1/2}^\natural,\\
h&=\frac{(\lambda|\lambda+2\rho_0)}{2(k+h^\lor)}+
\text{sdim}\,\fg_{1/2}\left(\frac{1}{16}+\frac{k}{8(k+h^\lor)}
\right)
-\frac{(\lambda|\theta)\left(k+h^\lor-1\right)}{2(k+h^\lor)},
\end{aligned}
\end{align}
where ${\rho_{1/2}}$ and $\rho_0$ are the ``rho'' vectors for
$\Delta_{1/2}^+$ and $\Delta_0^+$ respectively:
\begin{align}
                      \label{rho 1/2}
\rho_{1/2} &=\frac{1}{2}\sum_{\alpha \in \Delta_{1/2}^+}
(-1)^{p_\alpha} \alpha ,\\
                      \label{rho 0}
\rho_0&=\frac{1}{2}\sum_{\alpha \in \Delta_{0}^+}
(-1)^{p_\alpha} \alpha=\rho^\natural .
\end{align}

We introduce a contravariant bilinear form $B(\ket{a},\ket{b})$ (with
respect to the anti-involution (\ref{anti-involution}))
on the W-algebra Verma module
$M_k(\Lambda,h)$ with highest weight vector $\ket{\Lambda,h}$.
Contravariance means that
$B(J \ket{a},\ket{b})=B(\ket{a},\omega(J) \ket{b})$, for $J\in W_k(\fg,x)$
and $\ket{a},\ket{b} \in M_k(\Lambda,h)$.
The form is normalized by $B(\ket{\Lambda,h},\ket{\Lambda,h})=1$.
The kernel of the contravariant form $B$ is a maximal submodule of
$M_k(\Lambda,h)$. In the next section we use this property to compute
the determinant of this form.


\section{Determinant formula for minimal W-algebras}

\label{Determinant formulae for minimal W-algebras}

\setcounter{equation}{0}


This section contains the main result of the current paper:
the determinant formula for minimal W-algebras in the twisted case.
First we state the
result:
the determinant formula for the NS sector
(untwisted case) is presented in
Section~\ref{NS sector},
it was obtained by Kac and Wakimoto in \cite{KW1}
and is also derived in Section~\ref{Determinant formula}.
Section~\ref{Ramond sector} contains
the determinant formula for the Ramond
sector, in which modes of all the fields are integer.
We also include the lengthy formula for the general twisted case in
Section~\ref{General twisted sector}.
In Section~\ref{Determinant formula} we prove these determinant formulas.

Before we proceed to the determinant expressions
we would like to remind to the reader some notation:
$h$ is an eigenvalue of the Virasoro field zero mode $L_0$ on the
highest weight vector,
$\Lambda(v)$ is the eigenvalue of the dimension 1 fields
$J_0^{\{v\}}, v \in \fh^\natural$.
The number $r=\text{rank}\,\fg$.
$\rho_0$ and $\rho_{1/2}$
are the ``rho'' vectors for $\Delta_0^+$ and $\Delta_{1/2}^+$
respectively, they are given by (\ref{rho 1/2}) and (\ref{rho 0}).
The partition function $P_W(\widehat\tau)$ is a number
of partitions of $\widehat\tau$ to a sum of positive roots
of the W-algebra, where as usual odd roots appear maximum
one time in the sum.
The function $q(\alpha,n)$ in the degrees is
\begin{equation}
q(\alpha,n)=(-1)^{p_\alpha (n+1)},
\end{equation}
i.e.~it is equal to $-1$ if $\alpha$ is odd and $n$ is even,
and equal to 1 otherwise.

We would like to stress here that
the determinants are polynomials in
$\Lambda$ and $h$. If $q(\alpha,n)=-1$ then
the degree of the corresponding factor is negative;
it means that the factor cancels some other factor as it is explained below.
(And exactly as it happens in the twisted loop algebra determinant formula
(\ref{explicit affine det formula})).
Odd roots of a lie superalgebra are of two types:
isotropic ($(\alpha|\alpha)=0$) or
a half of an even root.
If the root $\alpha$ is isotropic, then the corresponding factor
${\mathcal N}_{n,m}^{\ \alpha \, }$ does not depend on $n$ and
the product over $n$ can be evaluated explicitly:
\begin{equation}
\prod_{n\in \mathbb{N}}{({\mathcal N}_{n,m}^{\ \alpha \, })}^{q(\alpha,n)
P_W(\widehat{\eta}- n(\alpha^\natural ,m))}
=
{({\mathcal N}_{1,m}^{\ \alpha \, })}
^{P_W^{\widehat\alpha}(\widehat{\eta}- (\alpha^\natural ,m))},
\end{equation}
where $\widehat\alpha=(\alpha^\natural,m)$ and
$P_W^{\widehat\alpha}(\widehat \tau)$ is the number of partitions
of $\widehat \tau$ to the sum of the W-algebra positive roots not including
the root $\widehat\alpha$ itself.

If the root $\alpha$ is a half of an even root, then the
factor ${\mathcal N}_{n,m}^{\ \alpha \, }$ cancels one of the
factors corresponding to the root $2\alpha$.
For example, if there is a root $\theta/2$ then one can express
\begin{equation}   \label{half_root_cancelation}
\begin{gathered}
\prod_{m,n \in \mathbb{N}} \! {\mathcal N}_{n,m}^{\ \theta \,}(k,\Lambda,h)^
{P_W (\widehat{\eta}-(0, mn))}
\hspace{-15pt}
\prod_{\scriptstyle n \in \mathbb{N},
\atop
\scriptstyle
m \in \frac{1}{2}+\e_{\theta/2}+\nz}
\hspace{-15pt}
{\mathcal{N}}_{n,m}^{\ \theta/2 \,} (k,\Lambda,h)^
{q(\theta/2,n)
P_W
(\widehat{\eta}- (0 ,m n))}=\\
= \hspace{-7pt}
\prod_{\scriptstyle
m,n \in \mathbb{N},
\atop
\scriptstyle
m-n \in 2\mathbb{Z}+2\e_{\theta/2}}
\hspace{-7pt}
\nu_{n,m}(k,\Lambda,h)^
{P_W(\widehat{\eta}- (0 ,\frac{m n}{2}))},
\end{gathered}
\end{equation}
where $\e_{\theta/2}=0$ in the NS case and $\e_{\theta/2}=1/2$
in the Ramond case, and
\begin{equation}
\nu_{n,m}(k,\Lambda,h)\equiv
{\mathcal N}_{\frac{n}{2},m}^{\ \theta \,}(k,\Lambda,h)=
{\mathcal N}_{n,\frac{m}{2}}^{\ \theta/2 \,}(k,\Lambda,h).
\end{equation}

Taking into account all the above remarks we proceed to the determinant
formulae.


\subsection{NS sector}

\label{NS sector}


Let $\widehat{\eta}=(\eta,s), \eta \in \fh^{\natural *},s \in
\frac{1}{2}\mathbb{Z}$. The determinant
${\det}^\text{NS}_{\widehat{\eta}}(k,\Lambda,h)$ of the Verma module with
highest weight $(\Lambda,h)$ on the weight space ${(\Lambda-\eta,h+s)}$
is given by
\begin{equation}
                            \label{NS formula}
\begin{aligned}
 {\det}^\text{NS}_{\widehat{\eta}}(k,\Lambda,h)&=
(k+h^\lor)^
{(r-1)\sum_{m,n \in \mathbb{N}} P_W (\widehat{\eta}-(0,m n))}
\!\prod_{m,n \in \mathbb{N}} \! {\mathcal N}_{n,m}^{\ \theta \,}(k,\Lambda,h)^
{P_W (\widehat{\eta}-(0, mn))}\\
&\times \!\!\!
\prod_{\scriptstyle
\beta \in \Delta_0
\atop
\scriptstyle
m,n \in \mathbb{N}}
\!\!
{\mathcal N}_{n,m}^{\ \beta \,} (k,\Lambda)^
{q(\beta,n)
P_W(\widehat{\eta}-n(\beta ,m))}
\!\!
\prod_{\scriptstyle
\beta \in \Delta_0^+
\atop
\scriptstyle
n \in \mathbb{N}}
\!\!
{\mathcal N}_{n,0}^{\ \beta \,} (k,\Lambda)^
{q(\beta,n)
P_W(\widehat{\eta}-n(\beta ,0))}\\
&\times \!\!\!
\prod_{\scriptstyle
\alpha \in \Delta_{1/2}
\atop
\scriptstyle
m \in \frac{1}{2}+\nz, n \in \mathbb{N}}
\!\!\!\!
{\mathcal{N}}_{n,m}^{\ \alpha \,} (k,\Lambda,h)^
{q(\alpha,n)
P_W
(\widehat{\eta}- n(\alpha^\natural ,m))},
\end{aligned}
\end{equation}
where
\begin{align}
{\mathcal N}_{n,m}^{\ \beta\,} (k,\Lambda)=
&
(\Lambda+\rho_0 |\beta)+m(k+h^\lor)-\tfrac{n}{2}(\beta|\beta),
\qquad
\beta \in \Delta_0,
\\
{\mathcal{N}}_{n,m}^{\ \alpha \,} (k,\Lambda,h)=
&h-\frac{1}{4(k+h^\lor)}\Big(
\big(2(\Lambda+\rho_0|\alpha^\natural)+2 m (k+h^\lor)-n(\alpha|\alpha)
\big)^2  \nonumber \\
&\phantom{h}
+2(\Lambda | \Lambda+2\rho_0)-
(k+1)^2 \Big),
\qquad
\alpha \in \Delta_{1/2},
\\
{\mathcal{N}}_{n,m}^{\ \theta \,}(k,\Lambda,h)=
&
h-\frac{1}{4(k+h^\lor)}\Big(
\big( m (k+h^\lor)-n
\big)^2
+2(\Lambda |\Lambda+2\rho_0)-
(k+1)^2 \Big).
\end{align}

The partition function
$P_W(\eta)$ is defined with respect to the
root system $\Delta_W$ of the untwisted (NS) sector
of the vertex algebra $W_k(\fg,x)$.
The set of positive roots $\Delta_W^+$ is
a disjoint union of
\begin{equation}
\begin{array}{c}
\{(\alpha,m) | \alpha \in \Delta_0, m \in \mathbb{N}\},
\{(\alpha,0) | \alpha \in \Delta_0^+\},\\
\{(\alpha^\natural,m) | \alpha \in \Delta_{1/2},
m \in \frac{1}{2}+\nz\},
\{(0,m) | m \in \mathbb{N}\},
\end{array}
\end{equation}
where
the multiplicity of the last set is $r=\text{rank}\,\fg$.


\subsection{Ramond sector}

\label{Ramond sector}


Let $\widehat{\eta}=(\eta,s),
\eta \in \fh^{\natural *},s \in \mathbb{Z}$.
The determinant
${\det}_{\widehat{\eta}}^{\text{R}}(k,\Lambda,h)$
of the Verma module with highest weight $(\Lambda,h)$
on the weight space ${(\Lambda-\eta,h+s)}$
is given by
\begin{equation}
                            \label{Ramond formula}
\begin{aligned}
 &{\det}_{\widehat{\eta}}^{\text{R}}(k,\Lambda,h)=
(k+h^\lor)^
{(r-1)\sum_{m,n \in \mathbb{N}} P_W (\widehat{\eta}-(0,m n))}
\!\prod_{m,n \in \mathbb{N}} \! {\mathcal{N}}_{n,m}^{\ \theta \, }(k,\Lambda,h)^
{P_W (\widehat{\eta}-(0, mn))}\\
&\times \!\!\!
\prod_{\scriptstyle
\beta \in \Delta_0
\atop
\scriptstyle
m,n \in \mathbb{N}}
\!\!
{\mathcal N}_{n,m}^{\ \beta \,} (k,\Lambda)^
{q(\beta,n)
P_W
(\widehat{\eta}-n(\beta ,m))}
\!\!
\prod_{\scriptstyle
\beta \in \Delta_0^+
\atop
\scriptstyle
n \in \mathbb{N}}
\!\!
{\mathcal N}_{n,0}^{\ \beta\, } (k,\Lambda)^
{q(\beta,n)
P_W
(\widehat{\eta}-n(\beta ,m))}\\
&\times \!\!\!
\prod_{\scriptstyle
\alpha \in \Delta_{1/2}
\atop
\scriptstyle
m,n \in \mathbb{N}}
\!\!
{\mathcal{N}}_{n,m}^{\ \alpha\, } (k,\Lambda,h)^
{q(\alpha,n)
P_W
(\widehat{\eta}- n(\alpha^\natural ,m))}
\!\!
\prod_{\scriptstyle
\alpha \in \Delta_{1/2}^+
\atop
\scriptstyle
 n \in \mathbb{N}}
\!\!
{\mathcal{N}}_{n,0}^{\ \alpha\, } (k,\Lambda,h)^
{q(\alpha,n)
P_W
(\widehat{\eta}- n (\alpha^\natural ,0))},
\end{aligned}
\end{equation}
where
\begin{align}
                                        \label{Ramond a(x)=0}
{\mathcal N}_{n,m}^{\ \beta\, } (k,\Lambda)=
&
(\lambda^\natural+\rho_0 |\beta)+m(k+h^\lor)-\tfrac{n}{2}(\beta|\beta),
\qquad
\beta \in \Delta_0,
\\
{\mathcal{N}}_{n,m}^{\ \alpha \,} (k,\Lambda,h)=
&h-\frac{1}{4(k+h^\lor)}\Big(
\big(2(\lambda^\natural+\rho_0|\alpha^\natural)+2 m (k+h^\lor)-n(\alpha|\alpha)
\big)^2  \nonumber \\
&\phantom{h}
+2(\lambda^\natural | \lambda^\natural+2\rho_0)-
(k+1)^2 \Big)+\frac{h^\lor-2}{8}\, ,
\qquad
\alpha \in \Delta_{1/2},
\\
%
%
{\mathcal{N}}_{n,m}^{\ \theta \,}(k,\Lambda,h)=
&
h-\frac{1}{4(k+h^\lor)}\Big(
\big( m (k+h^\lor)-n
\big)^2
+2(\lambda^\natural |\lambda^\natural+2\rho_0)
\nonumber \\
&\phantom{h}-
(k+1)^2 \Big)+\frac{h^\lor-2}{8}\, .
%
\end{align}
$\lambda^\natural$ in these formulas stands for
\begin{equation}
\lambda^\natural=\Lambda+\rho_{1/2}^\natural\, .
\end{equation}

The partition function
$P_W(\eta)$ is defined with respect to the
root system $\Delta_W$ of the Ramond sector
of the vertex algebra $W_k(\fg,x)$.
The set of positive roots $\Delta_W^+$ is
a disjoint union of
\begin{equation}   \label{Delta_W^+}
\begin{array}{c}
\{(\alpha,m) | \alpha \in \Delta_0, m \in \mathbb{N}\},
\{(\alpha,0) | \alpha \in \Delta_0^+\},\\
\{(\alpha^\natural,m) | \alpha \in \Delta_{1/2}, m \in \mathbb{N}\},
\{(\alpha^\natural,0) | \alpha \in \Delta_{1/2}^+\},\\
\{(0,m) | m \in \mathbb{N}\},
\end{array}
\end{equation}
where
the multiplicity of the last set is $r$.


\subsection{General twisted sector}

\label{General twisted sector}


Let $\widehat{\eta}=(\eta,s),
\eta \in \fh^{\natural *},s \in \mathbb{R}$.
The determinant
${\det}_{\widehat{\eta}}^{\text{tw}}(k,\Lambda,h)$
of the Verma module with highest weight $(\Lambda,h)$
on the weight space ${(\Lambda-\eta,h+s)}$
is given by
\begin{equation}
                            \label{twisted formula}
\begin{aligned}
 &{\det}_{\widehat{\eta}}^{\text{tw}}(k,\Lambda,h)=
(k+h^\lor)^
{(r-1)\sum_{m,n \in \mathbb{N}} P_W (\widehat{\eta}-(0,m n))}
\!\prod_{m,n \in \mathbb{N}} \! {\mathcal{N}}_{n,m}^{\ \theta \, }(k,\Lambda,h)^
{P_W (\widehat{\eta}-(0, mn))}\\
&\times
\hspace{-7pt}
\prod_{\scriptstyle
\beta \in \Delta_0^+
\atop
{\scriptstyle
m \in -\e_\beta+\mathbb{Z},
\atop
\scriptstyle
m>0,
n \in \mathbb{N}}}
\hspace{-7pt}
{\mathcal N}_{n,m}^{\, -\beta \,} (k,\Lambda)^
{q(\beta,n)
P_W(\widehat{\eta}-n(-\beta ,m))}
\hspace{-7pt}
\prod_{\scriptstyle
\beta \in \Delta_0^+
\atop
{\scriptstyle
m \in \e_\beta+\mathbb{Z},
\atop
\scriptstyle
m \ge 0,
n \in \mathbb{N}}}
\hspace{-7pt}
{\mathcal N}_{n,m}^{\ \beta\, } (k,\Lambda)^
{q(\beta,n)
P_W(\widehat{\eta}-n(\beta ,m))}\\
&\times \hspace{-12pt}
\prod_{\scriptstyle
\alpha \in \Delta_{1/2} \backslash \Delta_{1/2}^+
\atop
{\scriptstyle
m \in \e_\alpha+1/2+\mathbb{Z},
\atop
{\scriptstyle
m>0,
n \in \mathbb{N}}}}
\hspace{-12pt}
{\mathcal{N}}_{n,m}^{\ \alpha\, } (k,\Lambda,h)^{
q(\alpha,n)
P_W(\widehat{\eta}- n(\alpha^\natural ,m))}
\hspace{-15pt}
\prod_{\scriptstyle
\alpha \in \Delta_{1/2}^+
\atop
{\scriptstyle
m\in \e_\alpha+1/2+\mathbb{Z},
\atop
\scriptstyle
m \ge 0,
 n \in \mathbb{N}}}
\hspace{-15pt}
{\mathcal{N}}_{n,m}^{\ \alpha\, } (k,\Lambda,h)^{
q(\alpha,n)
P_W(\widehat{\eta}- n (\alpha^\natural ,0))},
\end{aligned}
\end{equation}
where
\begin{align}
{\mathcal N}_{n,m}^{\ \beta\, } (k,\Lambda)=
&
(\lambda^\natural+\widetilde\rho^\natural |\beta)+m(k+h^\lor)-\tfrac{n}{2}(\beta|\beta),
\qquad
\beta \in \Delta_0,
\\
{\mathcal{N}}_{n,m}^{\ \alpha\, } (k,\Lambda,h)=
&h-\frac{1}{4(k+h^\lor)}\Big(
\big(2(\lambda^\natural+\widetilde\rho^\natural|\alpha^\natural)+2 m (k+h^\lor)-n(\alpha|\alpha)
\big)^2
\nonumber \\
&
+2(\lambda^\natural | \lambda^\natural+2\widetilde\rho^\natural)
-\big(k+h^\lor-(\widetilde\rho|\theta)
\big)^2
 \nonumber \\
&
+2 k\sum_{\gamma\in\Delta_+}
{(-1)^{p_\gamma}\e_\gamma(1-\e_\gamma)}
\Big)
-h^\text{ch}-h^\text{ne},
\qquad
\alpha \in \Delta_{1/2},
\\
{\mathcal{N}}_{n,m}^{\ \theta\, }(k,\Lambda,h)=
&
h-\frac{1}{4(k+h^\lor)}\Big(
\big( m (k+h^\lor)-n
\big)^2
+2(\lambda^\natural | \lambda^\natural+2\widetilde\rho^\natural)
\nonumber\\
&
-\big(k+h^\lor-(\widetilde\rho|\theta)
\big)^2+2 k\sum_{\gamma\in\Delta_+}
{(-1)^{p_\gamma}\e_\gamma(1-\e_\gamma)}
\Big)
-h^\text{ch}-h^\text{ne}.
\end{align}
The twistings are assumed to be in the range $0\le\e_\alpha<1$
for all $\alpha\in\Delta_+$, the numbers $h^\text{ch}$ and
$h^\text{ne}$ are given in (\ref{h_ch}) and (\ref{h ne})
respectively,
$\lambda^\natural$ in these formulas stands for
\begin{equation}
\lambda^\natural=\Lambda
+\frac{1}{2} \sum_{\alpha \in \Delta_{1/2}}
(-1)^{p_\alpha} \alpha (\e_\alpha+\varkappa_\alpha),
\end{equation}
where $\varkappa_\alpha$ is defined in (\ref{varkappa}).

The function
$P_W(\eta)$ is the partition function
of the set of positive roots $\Delta_W^+$
of the W-algebra $W_k(\fg,x)$ in the twisted sector.
$\Delta_W^+$ is
a disjoint union of
\begin{equation}
\begin{array}{c}
\{(\alpha,m) | \alpha \in \Delta_0^+, m \in \e_\alpha +\mathbb{Z},m\ge0 \},
\{(-\alpha,m) | \alpha \in \Delta_0^+, m\in -\e_\alpha+\mathbb{Z},m>0\},\\
\{(\alpha^\natural,m) | \alpha \in \Delta_{1/2}^+,
m \in \e_\alpha+1/2+\mathbb{Z}, m \ge 0\},\\
\{(\alpha^\natural,m) | \alpha \in \Delta_+ \backslash \Delta_{1/2}^+,
m\in \e_\alpha+1/2+\mathbb{Z},m>0\},
\{(0,m) | m \in \mathbb{N}\},
\end{array}
\end{equation}
where again
the multiplicity of the last set is $r$.


\subsection{Derivation of the determinant formula}

\label{Determinant formula}



We prove here the determinant formula for minimal W-algebras, stated in
Sections~\ref{NS sector}, \ref{Ramond sector}, \ref{General twisted sector}.
To avoid very long expressions in the derivation we will do
part of the calculations for two most important cases: the NS (untwisted)
sector~(\ref{NS sector minimal})
and the Ramond sector~(\ref{Ramond sector minimal}).
The determinant formula in the untwisted case was obtained in~\cite{KW1},
we include its derivation here for completeness.
The NS and Ramond sectors are special cases of the general twisting
(see Section~\ref{Twist gluing}).
The computations in the general twisted case
(as anywhere in this paper the case $\e(\fh)=0$ only is
discussed) are similar to the those presented in this section.



The factors of the W-algebra $W(\fg, x)$ determinant formula are generically the same
as the factors of the underlying twisted loop algebra $\twg$ determinant formula,
the factors just have to be reexpressed in terms of the W-algebra weights.
The determinant formula vanishes if and only if there is a singular
vector in the Verma module. We recall from Section~\ref{Representation theory}
that if there is a singular vector $\ket{\widehat s}$ in the
highest weight module $R_{\widehat \lambda}$ of the affine vertex algebra
$V_k(\fg)$ then there is a singular vector $\ket{s}_k$ in the corresponding
highest weight module $Q_k(\lambda)$ of the vertex algebra
${\cal{C}}(\fg,x,k)=V_k(\fg)
\otimes F(A_{\text{ch}})
\otimes F(A_{\text{ne}})$
and it is given by
$\ket{s}_k=\ket{\widehat s}\times \ket{0}_\text{ch} \times \ket{0}_\text{ne}$.
The vector $\ket{s}_k$ is $d_0$-closed and if in addition it is not
$d_0$-exact then it is a singular vector in the W-algebra module
$M_k(\lambda)=H_0(Q_k(\lambda),d_0)$. We will see that generically
the vector $\ket{s}_k$ is not $d_0$-exact apart from a few cases listed
in the Corollary 
below. It also comes out that all the W-algebra singular vectors are given
by the above construction. It is proved 
by the standard degree counting, which shows that there is no room for other factors
in the determinant formula.


First we reexpress the factors of the affine determinant formula
(\ref{affine det formula})
in terms of the W-algebra weights.
Substituting $\widehat \rho=(\widetilde \rho,h^\lor,0)$ into
(\ref{affine det formula}) one gets the factors
\begin{equation}
\varphi^{\ \alpha\, }_{n,m}(\lambda,k)=(\lambda+\widetilde \rho|\alpha)+m
(k+h^\lor)-\tfrac{n}{2}(\alpha|\alpha),
\end{equation}
where $\widehat \alpha =(\alpha,0,m)$ is a positive root of the twisted loop
algebra $\twg$.
The ``twisted rho'' $\widetilde \rho$ is defined in (\ref{rho twisted}).
In the NS and Ramond cases it is equal to
\begin{align}
\text{NS: }& \widetilde \rho = \rho=\rho_0+\tfrac{1}{2}(h^\lor-1)
\theta,\\
\text{Ramond: }&\widetilde \rho=
\rho_0+\tfrac{1}{2}\theta.
\end{align}
In the untwisted case $\widetilde \rho=\rho$ in agreement with a
well known fact that the Weyl vector of affine Lie superalgebra
is equal to $\widehat\rho=(\rho,h^\lor,0)$. We will show now that
the Weyl vector $\widehat \rho=(\widetilde \rho,h^\lor,0)$
of a twisted loop algebra in the Ramond sector satisfies the set
of equations (\ref{rho def}), proving the conjecture of
Section~\ref{Affine vertex algebra} for the special case of Ramond twisting.
The simple roots of $\twg$ are
\begin{equation}
\begin{aligned}
\widehat \alpha_s &=(\alpha_s,0,0),\\
\widehat \beta_s &=(\beta_s-\theta,0,1/2),\\
\widehat \theta &= (\theta,0,0),
\end{aligned}
\end{equation}
where $\alpha_s$ and $\beta_s$ are simple roots of $\fg$,
such that $(\alpha_s|\theta)=0$ and $(\beta_s|\theta)=1/2$.
Compute the products of $\widehat \rho=(\widetilde \rho,h^\lor,0)$,
where $\widetilde \rho=\rho_0+\tfrac{1}{2}\theta$,
with our simple roots:
$(\widehat \rho|\widehat \alpha_s)=(\rho_0|\alpha_s)=
\tfrac{1}{2}(\alpha_s|\alpha_s)$ since $\rho_0$ is the ``rho''
vector for the set of roots $\Delta_0^+$;
$(\widehat \rho|\widehat \beta_s)
=(\widetilde \rho|\beta_s)-1+\tfrac{h^\lor}{2}=
(\rho|\beta_s)=\tfrac{1}{2}(\beta_s|\beta_s)=
\tfrac{1}{2}(\widehat\beta_s|\widehat\beta_s)$
since $\rho=\widetilde \rho+\tfrac{1}{2}(h^\lor-2)\theta$;
and $(\widehat \rho|\widehat \theta)=1$ is trivial.

The factor $\varphi^{\ \alpha}_{n,m}(\lambda,k)$ vanishes 
if and only if there is a singular vector in the $\twg$ module which appears first time
on the weight space $\widehat \lambda-\widehat \eta$,
where $\widehat \eta=(n \alpha,0, n m)$.
Then (with exception of the cases listed in the Lemma 
below)
there is also a singular vector in the $W_k(\fg,x)$ module
with weights $(\Lambda(\lambda,k),h(\lambda,k))$ appearing first time
on the weight space
$\left(\Lambda-n \alpha^\natural, h+n \left(
m+
\tfrac{1}{2}(\alpha|\theta)\right) \right)$.

Let us first discuss the case $(\alpha|\theta)=0$, which happens
when $\alpha\in \Delta_0$ or $\alpha=0$. Then the factor
is expressed as
\begin{equation}    \label{factor 0}
\varphi^{ \ \alpha}_{n, m}(\lambda,k)=(\lambda^\natural+\widetilde \rho^\natural|\alpha)
+m(k+h^\lor)-\tfrac{n}{2}(\alpha|\alpha), \quad \alpha(x)=0.
\end{equation}
Substituting $\lambda^\natural$ by a shifted $\Lambda$
using~(\ref{Lambda lambda}) one gets exactly
 the factor entering to the W-algebra determinant
formula. The factor doesn't depend on $h$.
The corresponding singular vector appears on the weight
space $(\Lambda-n \alpha,h+nm)$.

Now we proceed to the case $(\alpha|\theta) \ne 0$.
We would like to collect the factors which give rise to the
W-algebra module
singular vectors on the weight space $(\Lambda-n \alpha^\natural,h+nm)$.
These are two factors $\varphi^{\ \alpha}_{n, m-\frac{1}{2}(\alpha|\theta)}$ and
$\varphi^{\ \bar  \alpha}_{n, m+\frac{1}{2}(\alpha|\theta)}$, where $\bar \alpha$
is a ``mirror'' of the root $\alpha$:
\begin{equation}
\bar \alpha=\alpha^\natural-\tfrac{1}{2}(\alpha|\theta) \theta .
\end{equation}
So the following
expression is the W-algebra determinant factor for the singular vectors
on the weight
space $(\Lambda-n \alpha^\natural,h+nm)$:
\begin{equation}               \label{mathcal N}
\begin{aligned}
{\mathcal N}^{\ \alpha}_{n,m} &\equiv -\frac{1}{(k+h^\lor)(\alpha|\theta)^2}\,
\varphi^{\ \alpha}_{n, m-\tfrac{1}{2}(\alpha|\theta)}
\varphi^{\ \bar \alpha}_{n, m+\tfrac{1}{2}(\alpha|\theta)}=\\
&=\frac{1}{(k+h^\lor)(\alpha|\theta)^2}
\left(
\tfrac{1}{4}(\alpha|\theta)^2
\left((\lambda+\widetilde\rho|\theta)-k-h^\lor \right)^2-\right.\\
&-\left.
\left((\lambda^\natural+\widetilde\rho^\natural|\alpha^\natural)+
m(k+h^\lor)-\tfrac{n}{2}(\alpha|\alpha) \right)^2
\right), \quad \alpha\in \Delta_>.
\end{aligned}
\end{equation}

Next one has to express ${\mathcal N}^{\ \alpha}_{n,m}$ 
in terms of $\Lambda$ and $h$.
For that reason we rewrite (\ref{Lambda h NS}) and
(\ref{Lambda h Ramond}) as
\begin{align}
\text{NS: }&
h=\frac{1}{4(k+h^\lor)}
\left(2(\lambda^\natural|\lambda^\natural+2\rho_0)+
\left((\lambda|\theta)-k-1 \right)^2-(k+1)^2
\right),\\
\text{Ramond: }&
\begin{aligned}
h&=\frac{1}{4(k+h^\lor)}
\Big(2(\lambda^\natural|\lambda^\natural+2\rho_0)+
\left((\lambda|\theta)-k-h^\lor+1 \right)^2-\\
&
-(k+1)^2
\Big)
-\frac{h^\lor-2}{8}\, ,
\end{aligned}
\end{align}
where in the Ramond sector formula we used the fact that
$\text{sdim}\,\fg_{1/2}=2h^\lor-4$ in the case of minimal
gradation. Note that the part including the $(\lambda|\theta)$
term is the same as in (\ref{mathcal N}). So one can rewrite the
determinant factor as
\begin{align}
                                     \label{factor NS}
\text{NS: }&
\begin{aligned}
{\mathcal N}^{\ \alpha\, }_{n,m}
&=h-\frac{1}{4(k+h^\lor)}
\left(
2(\lambda^\natural|\lambda^\natural+2\rho_0)
-(k+1)^2
\right.\\
&+\left.
\frac{4}{(\alpha|\theta)^2}
\left((\lambda^\natural+\rho_0|\alpha^\natural)+
m(k+h^\lor)-\tfrac{n}{2}(\alpha|\alpha) \right)^2
\right),
\end{aligned}\\
                                     \label{factor Ramond}
\text{Ramond: }&
\begin{aligned}
{\mathcal N}^{\ \alpha\, }_{n,m}
&=h-\frac{1}{4(k+h^\lor)}
\Big(
2(\lambda^\natural|\lambda^\natural+2\rho_0)
-(k+1)^2
\\
&
+\frac{4}{(\alpha|\theta)^2}
\left((\lambda^\natural+\rho_0|\alpha^\natural)+
m(k+h^\lor)-\tfrac{n}{2}(\alpha|\alpha) \right)^2
\Big)
+\frac{h^\lor-2}{8}
\, ,
\end{aligned}
\end{align}
where $\alpha\in \Delta_>$, $\lambda^\natural=\Lambda$ in the NS case
and $\lambda^\natural=\Lambda+\rho_{1/2}^\natural$ in the Ramond case.


Now we should check which of the singular vectors of type
$\ket{\widehat s}\times \ket{0}_\mathrm{ch} \times \ket{0}_\mathrm{ne}$
($\ket{\widehat s}$ is a singular vector in the affine vertex algebra
highest weight module $R_{\widehat \lambda}$)
are $d_0$-exact, and so trivial in the $d_0$-homology.

Let $\ket{\lambda}_k=\ket{\widehat \lambda}\times
\ket{0}_\mathrm{ch} \times \ket{0}_\mathrm{ne}$
be a highest weight vector
in the highest weight module $Q_k(\lambda)$ of the vertex algebra
${\cal{C}}(\fg,x,k)=V_k(\fg)
\otimes F(A_{\mathrm{ch}})
\otimes F(A_{\mathrm{ne}})$.
Define the space $\xi$ as a span
of vectors of the form
$u_{\theta,-n_1}u_{\theta,-n_2}\cdots u_{\theta,-n_l} \ket{\lambda}_k$,
$n_i \in \mathbb{N}$.

\begin{lemma*}
                             \label{lemma1}

In the module $Q_k(\lambda)$ all the 
vectors  of the form
$\ket{\widehat t}\times \ket{0}_\mathrm{ch} \times \ket{0}_\mathrm{ne}$,
where $\ket{\widehat t}$ is a vector in the affine vertex algebra module
$R_{\widehat \lambda}$,
which are $d_0$-exact or differ from the highest weight vector $\ket{\lambda}_k$
by a $d_0$-exact vector,
belong to the space $\xi$ or to the space $u_{\alpha,-1+\e_\alpha}\, \xi$,
where
$\alpha \in \Delta_{1/2}, \, 1/2<\e_\alpha<1$  or
$\alpha \in \Delta_{1/2}\backslash \Delta_{1/2}^-,\, \e_\alpha=1/2$.

\end{lemma*}
\begin{proof}
In the case of minimal gradation $d_0$ has the following form:
\begin{equation}
\begin{aligned}
d_0=&
\sum_{
\alpha \in \Delta_{1/2}
\atop
n \in \e_\alpha+1/2+\ZZ
}
(-1)^{p_\al} u_{\al,n-1/2} \ghb^\al_{-n}+
\sum_{n \in \ZZ} u_{\theta,n-1} \ghb^\theta_{-n}\\
&-
\half \hspace{-10pt}
\sum_{
\alpha \in \Delta_{1/2}^+ \atop
 {n_1 \in \e_{\alpha}+1/2+\ZZ \atop
 n_2 \in -\e_{\alpha}+1/2+\ZZ}
}
\hspace{-10pt}
\str_{{\theta -  \alpha}, {\alpha}}^{\, \theta \, }\,
\ghg_{\gamma,n_1+n_2} \ghb^{\theta -  \alpha}_{-n_2} \ghb^{\al}_{-n_1}
+ (f|u_\theta) \ghb^\theta_0 +
\sum_{
\alpha \in \Delta_{1/2}
\atop
n \in \e_\alpha+1/2+\ZZ
}
\psi_{\alpha,n} \ghb^\alpha_{-n} .
\end{aligned}
\end{equation}
One has to analyze the action of $d_0$ on the ghost number 1
vector $c^\alpha_{n} \ket{\eta}$, $\ket{\eta} \in \xi$.
The action of $d_0$ on other type ghost number 1 vectors obviously
can not give a vector of the form
$\ket{\widehat t}\times \ket{0}_\text{ch} \times \ket{0}_\text{ne}$.
One can easily check that
\begin{equation}
                                  \label{d0short}
d_0 \, \ghg_{\theta,n} \ket{\eta} =
\left( r_1  \, u_{\theta,n-1}+ r_2 \, \delta_{0,n}\right) \ket{\eta},
\qquad n \in -\nz
\end{equation}
($r_1$ and $r_2$ are nonzero constants),
proving that any vector in $\xi$ is in the same homology class
as zero or the highest weight vector $\ket{\lambda}_k$.
In the same way
\begin{equation}
                                 \label{d0long}
\begin{split}
d_0 \, \ghg_{\al,m} \ket{\eta} =
\Big( r_3  \, u_{\al,m-1/2}+
r_4  \, \psi_{\al,m}+
r_5
\hspace{-10pt}
\sum_{m_1 \in -\e_{\alpha}+1/2+\ZZ}
\hspace{-10pt}
\ghg_{\gamma,m+m_1} \ghb^{\theta -  \alpha}_{-m_1}
\Big)\ket{\eta},
\\ 
\al \in \Delta_{1/2}, \, m<1/2, \, m\in \e_\al+1/2+\ZZ ,
\end{split}
\end{equation}
where $r_3,r_4,r_5$ are nonzero constants.
Now we observe that the second and the third terms in (\ref{d0long}) vanish
or proportional to $\ket{\eta}$
if and only if $0<m<1/2$ or $m=0,\, \al \in \Delta_{1/2}\backslash \Delta_{1/2}^-$,
and by this complete the proof of the lemma.
\end{proof}
\begin{corollary*}
The singular vectors of the form
$\ket{\widehat s}\times \ket{0}_\mathrm{ch} \times \ket{0}_\mathrm{ne}$
which do not give rise to the singular vectors
in the W-algebra highest weight module
are $(u_{\theta,-1})^n \ket{\lambda}_k$
and $(u_{\alpha,-1+\e_\alpha})^n \ket{\lambda}_k$,
$\alpha \in \Delta_{1/2}\backslash \Delta_{1/2}^-,\, \e_\alpha=1/2$,
$n \in \mathbb{N}$.
\end{corollary*}
\begin{proof}
The direct analysis of the structure of the twisted loop algebra module singular vectors
shows that these are the only singular vectors which belong
to the spaces $\xi$ and $u_{\alpha,-1+\e_\alpha}\, \xi$
discussed in the Lemma above.
\end{proof}

Next we just collect all the factors eliminating those
corresponding to the singular vectors which belong to the spaces
listed in the Lemma.
This elimination affects the lower bound for the running
indices in the determinant formula. And at the end we get
the factors of the determinant formulas in
Sections~\ref{NS sector}, \ref{Ramond sector}, \ref{General twisted sector}.

The degrees of the factors should not be less than
the W-algebra partitions since the singular vector generates
a maximal submodule in the W-algebra highest weight module.
So one gets that the expressions
in~(\ref{NS formula}, \ref{Ramond formula}, \ref{twisted formula})
are divisors of the determinant formulas.
Now by using the usual counting arguments
coming from the estimation of the power
of the determinant formula
one shows that actually the powers in these expressions are
already equal to the maximal estimation and
therefore there is no room for other factors.
We will not reproduce here the degree counting,
since it is fully equivalent to the one performed
in \cite{KW1} for the untwisted case.



%



\section{Examples}

\label{Examples}

\setcounter{equation}{0}


In this section we discuss all the simple Lie superalgebras of rank
up to 2 except $sl(2)$: $sl(3), so(5), G_2, osp(1|2), sl(2|1),
osp(3|2), osp(1|4), psl(2|2)$. The $sl(2)$ Lie algebra is
not presented here since there is only untwisted quantum reduction on it.
The $osp(1|2)$ superalgebra is of rank 1, all the rest
in the list above are of rank 2.

In all the examples the quantum reduction
corresponding to the minimal gradation on the Lie superalgebra is discussed.
The minimal gradation is generated by the $sl(2)$ embedding associated to the highest
root $\theta$ as it is described in Section~\ref{Gradation on a Lie superalgebra}.

The generating element $x$ is normalized as $(x|x)=1/2$
(since $x\equiv \theta/2$ and $(\theta|\theta)=2$).
In the case of rank 2 algebra the orthogonal to $x$ Cartan
generator is denoted by $y$: $y \in \fh, (y|x)=(y|\theta)=0$.
We normalize it by $(y|y)=\pm 1/2$, where the sign should be
chosen according to the sign of the metric in the
corresponding root space direction:
$(y|y)=-1/2$ for $sl(2|1), osp(3|2), psl(2|2)$
Lie superalgebras and $(y|y)=1/2$ for the rest of the rank--2 examples.

We discuss only the case when the Cartan subalgebra is untwisted:
$\e(x)=\e(y)=0.$
One should note that in all the rank--2 examples there is yet another
possibility: $\e(y)=1/2$, which is not studied in the current
paper.

At the beginning of each example there is a figure showing
the root system of the Lie superalgebra under discussion.
Even and odd roots are shown by arrows of different style.


\subsection{$osp(1|2)$}


\begin{figure}[h]
\vspace{-5pt}
\centering
\includegraphics[width=200pt]{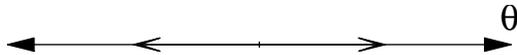}
\vspace{-5pt}
{\caption{$osp(1|2)$ root system.}  \label{fig osp12}}
\vspace{-5pt}
\end{figure}
%
The $osp(1|2)\approx B(0,1)$ algebra is the simplest
Lie superalgebra. It is of rank 1.
There is only one good gradation on it: the obvious Dynkin gradation.
The quantum reduction procedure on $osp(1|2)$ is described in \cite{KRW} and \cite{KW1}:
one gets the famous $N=1$ superconformal algebra,
an extension of the Virasoro algebra by a dimension-$3/2$ fermionic primary field.

The $osp(1|2)$ algebra has two positive roots: $\theta$ and
$\theta/2$.
The dual Coxeter number is $h^\lor=\frac{1}{2}\, \text{sdim}\,\fg_{1/2}+2=3/2$.
The central charge is given by (\ref{c_minimal_gradation}):
\begin{equation}
c=\frac{k}{k+3/2}-6k-5/2\, .
\end{equation}

There are two possible twistings: $\e\equiv \e_{\theta/2}=0$ leads
to half-integer modes of the dimension $3/2$ operator (the NS sector);
$\e=1/2$ leads to integer modes (the Ramond sector).

With a help of formula (\ref{half_root_cancelation})
one gets the well known \cite{Kac det, Friedan:1984rv, Meurman:1986gr}
determinant formula for
the $N=1$ superconformal algebra:
\begin{equation}
\text{det}_\eta(k,h)=\hspace{-7pt} \prod_{\scriptstyle m,n \in
\mathbb{N}, \atop \scriptstyle m-n \in 2\mathbb{Z}+2\e}
\hspace{-7pt} \nu_{n,m}(k,h)^ {P_W(\eta- \frac{m n}{2})},
\end{equation}
where
\begin{equation}
\nu_{n,m}(k,h)= h-\frac{1}{4(k+\frac{3}{2})}\bigg(
\big(m(k+\frac{3}{2})-\frac{n}{2}\big)^2-(k+1)^2
\bigg)-\frac{\e}{8}\, .
\end{equation}
The partition function of the $N=1$ superconformal algebra can be easily
expressed using the following generating function:
\begin{equation}
\prod_{l=1}^\infty
\frac{1 + x^{l - 1/2+\e}}{1 - x^l}=
\sum_{n\in(\frac{1}{2}+\e)\nz } P_W(n)x^n .
\end{equation}


\subsection{$sl(2|1)$}


\begin{figure}[h]
\vspace{-5pt}
\centering
\includegraphics[width=200pt]{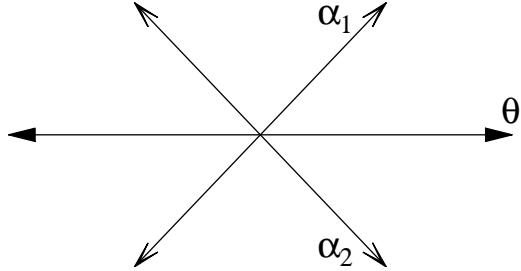}
\vspace{-5pt}
{\caption{$sl(2|1)$ root system.}  \label{fig sl21}}
\vspace{-5pt}
\end{figure}
%
The $sl(2|1) \approx A(1,0)$ algebra
is a rank--2 Lie superalgebra with 4 even and 4 odd generators.
The root system consists of one pair of even roots
($\theta, -\theta$) and two pairs of odd isotropic roots
($\alpha_1,-\alpha_1; \alpha_2,-\alpha_2$).
The system of positive roots, for which $\theta$ is a highest
root, is
$\Delta_+=\{ \alpha_1,\alpha_2,\theta=\alpha_1+\alpha_2 \}$.
The product between the simple roots is $(\alpha_1|\alpha_2)=1$.
The dual Coxeter number is $h^\lor=1$.

There is one Dynkin gradation (it is also the minimal one),
which corresponds to $f=u_{-\theta}$.
All other good gradations may be obtained from the
Dynkin one by changing the element $x$, which generates the
gradation.
With respect to the minimal gradation $\Delta_0=\varnothing,
\Delta_{1/2}=\{\alpha_1,\alpha_2\}, \Delta_{-1/2}=\{-\alpha_1,-\alpha_2\}$.

The W-algebra obtained from the minimal gradation is the $N=2$
superconformal algebra. It is generated by four fields:
the Virasoro field $L \sim J^{\{f\}}$, two dimension-$3/2$
fermionic fields $G^+ \sim J^{\{-\alpha_2\}}$
and $G^- \sim J^{\{-\alpha_1\}}$, and one dimension-1 bosonic current
$J = 2 J^{\{y\}}$, where $y$ is defined in the beginning of Section~\ref{Examples}.
One has to introduce ``2'' in the definition of the $U(1)$ current
$J$ to get it conventionally normalized: then $G^+$ and $G^-$
have $U(1)$ charges $+1$  and $-1$ respectively.
The explicit expressions for the fields can be found in
\cite{KRW}, Section 7. The central charge is
\begin{equation}
c=-6k-3.
\end{equation}

If $\e(\fh)=0$ then there is a one parameter family of twistings:
$\e_{\alpha_2} \equiv \e,\ \e_{\alpha_1}=1-\e$. The modes of the
bosonic fields $L$ and $J$ are integer, $n \in -\e+1/2 +\mathbb{Z}$
in $G^+_n$ and $m \in \e+1/2+\mathbb{Z}$ in $G^-_m$.
The untwisted case $\e=0$ leads to the NS sector, the case
$\e=1/2$ gives the Ramond sector. Different sectors (different $\e$)
are isomorphic to the NS sector, the isomorphism is given by the
so called $U(1)$ flow \cite{Schwimmer:1986mf}.

The separation of $\Delta_{1/2}$ and $\Delta_{-1/2}$ to positive
and negative parts (see Section~\ref{minimal representation theory}) is
made with respect to $y\in \fh^\natural$:
$\Delta_{1/2}^+=\{\alpha_1\}$, $\Delta_{-1/2}^+=\{-\alpha_2\}$,
$\Delta_{1/2}^-=\{\alpha_2\}$, $\Delta_{-1/2}^-=\{-\alpha_1\}$.
So in the Ramond sector $G^+_0$ annihilates the highest weight state
(while $G^-_0$ does not),
since $G^+$ is associated to the root $-\alpha_2 \in \Delta_{-1/2}^+$.
The ``rho'' vectors are $\rho_0=0$ and $\rho_{1/2}=-\frac{1}{2}\alpha_1$.

Although the determinant formula for the general twisting can be obtained from
the NS sector determinant formula using the $U(1)$ flow, we want to use
our general formulae from Sections~\ref{Ramond sector}, \ref{General twisted sector}.
For the NS and Ramond sectors
the determinant formula is
\begin{equation}                \label{det sl(2|1)}
\begin{aligned}
&\text{det}_{\widehat \eta}(k,q,h)=
(k+1)^
{\sum_{m,n \in \mathbb{N}} P_W (\widehat{\eta}-(0,m n))}
\!\prod_{m,n \in \mathbb{N}} \! {\mathcal{N}}_{n,m}^{\ \theta \, }(k,q,h)^
{P_W (\widehat{\eta}-(0, mn))}\\
& \times
\hspace{-5pt}
\prod_{m\in\frac{1}{2}-\e+\nz}
\hspace{-5pt}
{\mathcal N}_{1,m}^{\ \alpha_1}(k,q,h)^{P_W^{(1,m)}(\widehat{\eta}-(1, m))}
\hspace{-5pt}
\prod_{m\in\frac{1}{2}+\e+\nz}
\hspace{-5pt}
{\mathcal N}_{1,m}^{\ \alpha_2}(k,q,h)^{P_W^{(-1,m)}(\widehat{\eta}-(-1, m))},
\end{aligned}
\end{equation}
where
\begin{align}
{\mathcal N}_{n,m}^{\ \theta} (k,q,h)&=
h-\frac{1}{4(k+1)} \bigg(
(m(k+1)-n)^2-(q-\e)^2-(k+1)^2 \bigg)-\frac{\e^2}{2}\, ,\\
{\mathcal N}_{1,m}^{\, \alpha_1 (\alpha_2)} (k,q,h)&=           \label{last det sl(2|1)}
h-m\big(m(k+1) \mp (q-\e)\big)-\frac{\e^2}{2}+\frac{k+1}{4}\, ,
\end{align}
$q=2\Lambda$ is the $J_0$ eigenvalue, $\e=0 (1/2)$ in the NS (Ramond) sector.
This determinant formula was obtained in \cite{Boucher:1986bh} and in \cite{Kato:1986td}.

Applying the formulae in Section~\ref{General twisted sector}
to the $sl(2|1)$ Lie superalgebra
one gets the determinant formula of the $N=2$ algebra for the case of general
twisting $\e$. It is not surprising that after a simplification the determinant
formula becomes the same as in the case of NS and Ramond
sectors~(\ref{det sl(2|1)}-\ref{last det sl(2|1)}),
the only modification is that now $\e$ is a continuous parameter in the range
$-1/2<\e \le 1/2$.

The partitions are expressed as coefficients of the power expansion of the following
generating functions:
\begin{align}
\prod_{l=1}^\infty \frac{(1+x\, y^{l-1/2-\e})(1+x^{-1} y^{l-1/2+\e})}
{(1-y^l)^2}&=
\sum P_W(n_1,n_2)\, x^{n_1} y^{n_2}  ,\\
\frac{1}{1+x^{j_1} y^{j_2}}
\prod_{l=1}^\infty \frac{(1+x\, y^{l-1/2-\e})(1+x^{-1} y^{l-1/2+\e})}
{(1-y^l)^2}&=
\sum P_W^{(j_1,j_2)}(n_1,n_2)\, x^{n_1} y^{n_2}  .
\end{align}

The determinant formula for the case when the orthogonal
Cartan generator $y \in \fh^\natural$ is twisted ($\e(y)=1/2$) is calculated using
quantum reduction in \cite{KW2}.



\subsection{$sl(3)$}


\begin{figure}[h]
\vspace{-5pt}
\centering
\includegraphics[width=150pt]{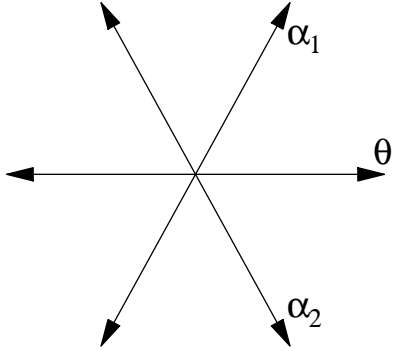}
\vspace{-5pt}
{\caption{$sl(3)$ root system.}  \label{fig sl3}}
\vspace{-5pt}
\end{figure}
%
The root system of the $sl(3)\approx A_2$ Lie algebra is shown on Figure~\ref{fig sl3}.
We use the normalization $(\alpha|\alpha)=2$,
the product of simple roots is $(\alpha_1|\alpha_2)=-1$, the dual Coxeter number
is $h^\lor=3$.
The quantum reduction on $sl(3)$ is very similar to the quantum reduction
on $sl(2|1)$, the difference comes from the fact that all the roots of $sl(3)$
are even. The quantum reduction of the minimal gradation on $sl(3)$ is described
in \cite{Bershadsky:1990bg}, it leads to the Bershadsky--Polyakov algebra.
The algebra resembles the $N=2$ superconformal algebra, but the dimension-$3/2$
generators $G^+$ and $G^-$ are bosonic, and their \ope\ contains non-linear terms
of type $\NO{J}{J}$.
The central charge of the algebra is
\begin{equation}
c=\frac{8\,k}{k+3}-6k-1\, .
\end{equation}
All the discussion of the previous subsection is applicable here.
There also exists a $U(1)$ flow. The determinant formula is
\begin{equation}                \label{det sl(3)}
\begin{aligned}
&\text{det}_{\widehat \eta}(k,q,h)=
(k+3)^
{\sum_{m,n \in \mathbb{N}} P_W (\widehat{\eta}-(0,m n))}
\!\prod_{m,n \in \mathbb{N}} \! {\mathcal{N}}_{n,m}^{\ \theta \, }(k,q,h)^
{P_W (\widehat{\eta}-(0, mn))}\\
& \times
\hspace{-5pt}
\prod_
{n\in \mathbb{N},
\atop
m\in\frac{1}{2}-\e+\nz}
\hspace{-5pt}
{\mathcal N}_{n,m}^{\ \alpha_1}(k,q,h)^{P_W(\widehat{\eta}-n(1, m))}
\hspace{-5pt}
\prod_{
n\in \mathbb{N},
\atop
m\in\frac{1}{2}+\e+\nz}
\hspace{-5pt}
{\mathcal N}_{n,m}^{\ \alpha_2}(k,q,h)^{P_W(\widehat{\eta}-n(-1, m))},
\end{aligned}
\end{equation}
where
\begin{align}
{\mathcal N}_{n,m}^{\ \theta} (k,q,h)&=
h-\frac{1}{4(k+3)} \bigg(
(m(k+3)-n)^2+3(q+\e)^2-(k+1)^2 \bigg)+\frac{\e^2}{2}\, ,\\
{\mathcal N}_{n,m}^{\, \alpha_1 (\alpha_2)} (k,q,h)&=
h-\frac{1}{4(k+3)} \bigg(
\big(2m(k+3)-2n \pm 3(q+\e) \big)^2 +\nonumber\\
& \phantom{h-\frac{1}{4(k+3)} \bigg(m+3)-n)^2+}
+3(q+\e)^2-(k+1)^2 \bigg)+\frac{\e^2}{2}
\, ,
\end{align}
$q=\frac{2}{\sqrt{3}}\Lambda$ is the $J_0$ eigenvalue, $\e \equiv \e_{\alpha_2}$
is taken in the range $-1/2 <\e \le 1/2$,
in particular $\e=0$ corresponds to the NS sector,
$\e=1/2$ -- to the Ramond sector.

The partition generating function is
\begin{equation}
\prod_{l=1}^\infty \frac{1}{(1-x\, y^{l-1/2-\e})(1-x^{-1} y^{l-1/2+\e})
(1-y^l)^2}=
\sum P_W(n_1,n_2)\, x^{n_1} y^{n_2} .
\end{equation}



\subsection{$osp(3|2)$}


\begin{figure}[h]
\vspace{-5pt}
\centering
\includegraphics[width=200pt]{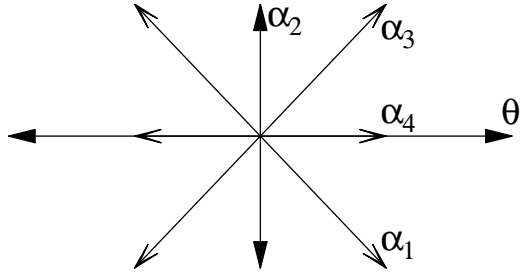}
\vspace{-5pt}
{\caption{$osp(3|2)$ root system.}  \label{fig osp32}}
\vspace{-5pt}
\end{figure}
%
The $osp(3|2)\approx B(1,1)$ Lie superalgebra has 6 even and 6 odd generators.
The root system is shown on Figure~\ref{fig osp32}.
There are 2 pairs of even roots $\theta,-\theta;\alpha_2,-\alpha_2$
and 3 pairs of odd roots
$\alpha_1,-\alpha_1;\alpha_3,-\alpha_3;\alpha_4,-\alpha_4$.
The products
between simple roots are $(\alpha_1|\alpha_1)=0$,
$(\alpha_2|\alpha_2)=-1/2$, $(\alpha_1|\alpha_2)=1/2$.
With respect to the minimal gradation $\Delta_0=\{\alpha_2,-\alpha_2\}$,
$\Delta_{1/2}=\{\alpha_3,\alpha_4,\alpha_1\}$.

The quantum hamiltonian reduction on the minimal gradation
of $osp(3|2)$ gives the $so(3)$ invariant superconformal algebra
of \cite{Miki:1989ri} (and references therein).
Besides the energy--momentum field,
there are three bosonic dimension-1 fields
generating an affine vertex algebra $V_{-4(k+1/2)}(sl(2))$
and three fermionic dimension-$3/2$ fields in the triplet representation
of the $sl(2)$:
\begin{equation}
\begin{aligned}
J^+ &\sim J^{\{\alpha_2\}}, \\
J&=J^{\{2y\}},\\
J^-&\sim J^{\{{-\alpha_2}\}},
\end{aligned}
\qquad
\begin{aligned}
G^+ &\sim J^{\{-\alpha_1\}}, \\
G&\sim J^{\{-\alpha_4\}},\\
G^-&\sim J^{\{{-\alpha_3}\}}.
\end{aligned}
\end{equation}
The explicit reduction formulas can be found in \cite{KW1}, Section 8.5.
For \ope s see \cite{Miki:1989ri}.
The central charge of the W-algebra is
\begin{equation}
c=-6\, k -7/2 .
\end{equation}

The twist numbers are parameterized by one discrete parameter
$\sigma \equiv \e_{\alpha_4}=0 \text{ or } 1/2$ and one continuous parameter
$\e \equiv \e_{\alpha_1}$. Then $\e_{\alpha_3}=1-\e$ and
$\e_{\alpha_2}=1-\e-\sigma$.
The NS sector is obtained if
$\e=\sigma=0$, the Ramond sector is given by $\e=\sigma=1/2$.
There is an isomorphism between different twisted sectors of the algebra
(an analogue of the $U(1)$ flow in the case of $N=2$ superconformal algebra):
\begin{equation}
\begin{aligned}
J^+_n &\mapsto J^+_{n-\e}\, ,\\
J^-_n &\mapsto J^-_{n+\e}\, ,
\end{aligned}
\quad
\begin{aligned}
G^+_n &\mapsto G^+_{n-\e}\, ,\\
G^-_n &\mapsto G^-_{n+\e}\, ,
\end{aligned}
\quad
\begin{aligned}
J_n &\mapsto J_{n}+\e(2k+1) \delta_{n,0}\, ,\\
L_n &\mapsto L_{n}-\e J_n-\e^2(k+\tfrac{1}{2})\delta_{n,0}\, ,\\
G_n &\mapsto G_{n}\, .
\end{aligned}
\end{equation}
The general twisted sector
of the W-algebra with $\sigma=0$ ($\sigma=1/2$)
is isomorphic to the NS sector (Ramond sector).
We state below only the NS and Ramond sectors determinant formulas,
the determinant
formula for the general twisted sector may be obtained from the
determinant formula for the NS or Ramond sector using this isomorphism.

Inserting the quantities $h^\lor=1/2$, $\rho_0=-1/2 \alpha_2$,
$\rho_{1/2}=-1/2 \alpha_3$  into the W-algebra determinant
formula one gets the following expression
\begin{equation}
\begin{aligned}
&\text{det}_{\widehat \eta}(k,q,h)=
(k+\tfrac{1}{2})^
{\sum_{m,n \in \mathbb{N}} P_W (\widehat{\eta}-(0,m n))}
\!\prod_{\scriptstyle m,n \in
\mathbb{N}, \atop \scriptstyle m-n \in 2\mathbb{Z}+2\sigma}
\! \nu_{n,m}(k,h)^ {P_W(\widehat \eta- (0,\frac{m n}{2}))}
\\
& \times
\hspace{-5pt}
\prod_
{
m\in\frac{1}{2}-\sigma+\nz}
\hspace{-5pt}
{\mathcal N}_{1,m}^{\ \alpha_3}(k,q,h)^{P_W^{(1,m)}(\widehat{\eta}-(1, m))}
\hspace{-5pt}
\prod_{
m\in\frac{1}{2}+\sigma+\nz}
\hspace{-5pt}
{\mathcal N}_{1,m}^{\ \alpha_1}(k,q,h)^{P_W^{(-1,m)}(\widehat{\eta}-(-1, m))}
\\
& \times
\!
\prod_
{n \in \mathbb{N},\,
m\in\nz}
\!
{\mathcal N}_{n,m}^{\ \alpha_2}(k,q)^{P_W(\widehat{\eta}-n(1, m))}
\!
\prod_{
m,n \in \mathbb{N}}
\!
{\mathcal N}_{n,m}^{\, -\alpha_2}(k,q)^{P_W(\widehat{\eta}-n(-1, m))},
\end{aligned}
\end{equation}
where
\begin{align}
\nu_{n,m}(k,q,h)&=h
+
\frac{k}{4}
+
\frac{3(1-\sigma)}{8}-
  \nonumber \\
&
-\frac{1}{4(k+\frac{1}{2})}
\bigg(
\big(m(k+\frac{1}{2})-\frac{n}{2}\big)^2-
(q+\frac{1}{2}-\sigma)^2
\bigg),
\\
{\mathcal N}_{1,m}^{ \alpha_{1}(\alpha_3)}
(k,q,h)&=
h-m^2(k+\frac{1}{2}) \mp m(q+\frac{1}{2}-\sigma)+
\frac{k}{4}
+
\frac{3(1-\sigma)}{8}\, ,\\
{\mathcal N}_{n,m}^{\pm \alpha_2}(k,q)&=
\mp \frac{q+\frac{1}{2}-\sigma}{2}+m(k+\frac{1}{2})+\frac{n}{4}\, ,
\end{align}
where $q=2\Lambda$ is the eigenvalue of $J_0$, and $\e=\sigma=0 \text{ or }1/2$.
The partitions are given by
\begin{align}
&\prod_{l=1}^\infty \frac{(1+x\, y^{l-1/2-\sigma})(1+x^{-1} y^{l-1/2+\sigma})
(1+ y^{l-1/2+\sigma})}
{(1-y^l)^2 (1-x y^{l-1})(1-x^{-1} y^{l})}
=
\sum P_W(n_1,n_2)\, x^{n_1} y^{n_2}  ,\\
&
\frac{1}{1+x^{j_1} y^{j_2}}
\prod_{l=1}^\infty \frac{(1+x\, y^{l-1/2-\sigma})(1+x^{-1} y^{l-1/2+\sigma})
(1+ y^{l-1/2+\sigma})}
{(1-y^l)^2 (1-x y^{l-1})(1-x^{-1} y^{l})}
=  \nonumber\\
&\phantom{aaaaaaaaaaaaaaaaaaaaaaaaaaaaaaaaaaaaaaaa} 
=\sum P_W^{(j_1,j_2)}(n_1,n_2)\, x^{n_1} y^{n_2}  .
\end{align}

The determinant formula was first obtained in \cite{Miki:1989ri}.



\subsection{$so(5)$}


\begin{figure}[h]
\vspace{-5pt}
\centering
\includegraphics[width=150pt]{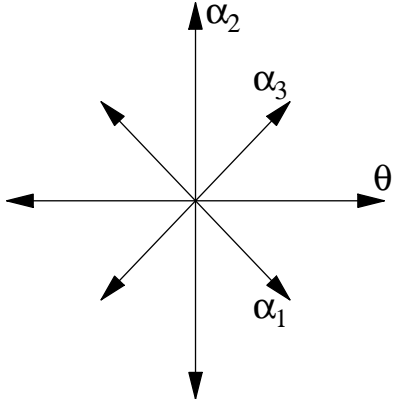}
\vspace{-5pt}
{\caption{$so(5)$ root system.}  \label{fig so5}}
\vspace{-5pt}
\end{figure}
%
The root system of $so(5) \approx B_2$
is shown on Figure~\ref{fig so5}.
There are two pairs of long roots and two pairs of short roots.
The simple roots are $\alpha_1$ and $\alpha_2$ with
product between them $(\alpha_1|\alpha_2)=-1$. Other positive roots are
$\alpha_3=\alpha_1+\alpha_2$ and $\theta=2\alpha_1+\alpha_2$.
With respect to the minimal gradation associated to $\theta$
one has $\Delta_0=\{\alpha_2, -\alpha_2\}, \Delta_{1/2}=\{\alpha_3, \alpha_1\}$.

The quantum reduction corresponding to the minimal gradation on
$so(5)$ gives a W-algebra which is generated by the Virasoro field,
three bosonic dimension--1 fields forming the $sl(2)$ affine vertex algebra
on the level $k_0=k+1/2$ ($V_{k+1/2}(sl(2))$),
and two bosonic dimension--$3/2$ fields in the doublet representation
of the $sl(2)$:
\begin{equation}
\begin{aligned}
J^+ &\sim J^{\{\alpha_2\}}, \\
J&=J^{\{y\}},\\
J^-&\sim J^{\{{-\alpha_2}\}},
\end{aligned}
\qquad
\begin{aligned}
G^+ &\sim J^{\{-\alpha_1\}}, \\
G^-&\sim J^{\{{-\alpha_3}\}}.
\end{aligned}
\end{equation}
The central charge of the W-algebra is
\begin{equation}
c=\frac{10\, k}{k+3}-6k-1.
\end{equation}

The twistings are parameterized by one continuous
parameter $\e \equiv \e_{\alpha_1}$. Other twist numbers
are related to $\e$ as $\e_{\alpha_3}=-\e$, $\e_{\alpha_2}=-2\e$.
The NS sector corresponds to $\e=0$, the Ramond sector is obtained
when $\e=1/2$. The isomorphism connecting different sectors is given by
\begin{equation}
\begin{aligned}
J^+_n &\mapsto J^+_{n-2\e}\, ,\\
J^-_n &\mapsto J^-_{n+2\e}\, ,
\end{aligned}
\quad
\begin{aligned}
G^+_n &\mapsto G^+_{n-\e}\, ,\\
G^-_n &\mapsto G^-_{n+\e}\, ,
\end{aligned}
\quad
\begin{aligned}
J_n &\mapsto J_{n}-\e(k+\tfrac{1}{2}) \delta_{n,0}\, ,\\
L_n &\mapsto L_{n}-2\e J_n+\e^2(k+\tfrac{1}{2})\delta_{n,0}\, .
\end{aligned}
\end{equation}

To write down the determinant formula for the W-algebra one uses
$h^\lor=3$, $\rho_0=\alpha_2/2$, $\rho_{1/2}^\natural=\alpha_3^\natural/2$.
The determinant formula for the NS and Ramond sector is
\begin{equation}                \label{det so(5)}
\begin{aligned}
&\text{det}_{\widehat \eta}(k,q,h)=
(k+3)^
{\sum_{m,n \in \mathbb{N}} P_W (\widehat{\eta}-(0,m n))}
\!\prod_{m,n \in \mathbb{N}} \! {\mathcal{N}}_{n,m}^{\ \theta \, }(k,q,h)^
{P_W (\widehat{\eta}-(0, mn))}\\
& \times
\hspace{-5pt}
\prod_
{n\in \mathbb{N},
\atop
m\in\frac{1}{2}+\e+\nz}
\hspace{-5pt}
{\mathcal N}_{n,m}^{\ \alpha_1}(k,q,h)^{P_W(\widehat{\eta}-n(-1/2, m))}
\hspace{-5pt}
\prod_{
n\in \mathbb{N},
\atop
m\in\frac{1}{2}-\e+\nz}
\hspace{-5pt}
{\mathcal N}_{n,m}^{\ \alpha_3}(k,q,h)^{P_W(\widehat{\eta}-n(1/2, m))}
\\
& \times
\!
\prod_
{n \in \mathbb{N},\,
m\in\nz}
\!
{\mathcal N}_{n,m}^{\ \alpha_2}(k,q)^{P_W(\widehat{\eta}-n(1, m))}
\!
\prod_{
m,n \in \mathbb{N}}
\!
{\mathcal N}_{n,m}^{\, -\alpha_2}(k,q)^{P_W(\widehat{\eta}-n(-1, m))},
\end{aligned}
\end{equation}
where
\begin{align}
{\mathcal N}_{n,m}^{\ \theta}(k,q,h)&=h
-\frac{1}{4(k+3)}
\bigg(
\big(m(k+3)-n \big)^2+
 \nonumber \\
&  \phantom{aaaaaaaaaaaaaaa\,\,}
+(2q+1+\e)^2-k^2-2k-2
\bigg)+\frac{\e^2}{2},
\\
{\mathcal N}_{n,m}^{ \alpha_{1}(\alpha_3)}
(k,q,h)&=
h
-\frac{1}{4(k+3)}
\bigg(
\big(\mp(2q+1+\e)+2m(k+3)-n \big)^2+ \nonumber \\
& \phantom{aaaaaaaaaaaaaaa\,\,}
+(2q+1+\e)^2-k^2-2k-2
\bigg)+\frac{\e^2}{2}
\, ,\\
{\mathcal N}_{n,m}^{\pm \alpha_2}(k,q)&=
\pm (2q+1+\e)+m(k+3)-n\, ,
\end{align}
here $q=\Lambda$ is the $J_0$ eigenvalue and $\e=0$ or $1/2$.
The partition function is
%
\begin{equation}
\begin{split}
\prod_{l=1}^\infty
\frac{1}
{(1-x^{1/2} y^{l-1/2-\e})
(1-y^l)^2 (1-x^{-1/2} y^{l-1/2+\e})}& \\
\times \frac{1}
{(1-x\,y^{l-1})(1-x^{-1}\,y^{l})}
&=
\sum P_W(n_1,n_2)\, x^{n_1} y^{n_2} .
\end{split}
\end{equation}


\subsection{$psl(2|2)$}


\begin{figure}[h]
\vspace{-5pt}
\centering
\includegraphics[width=150pt]{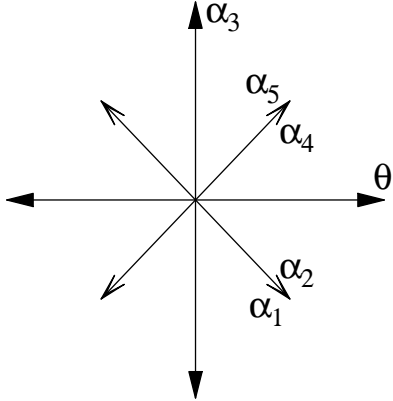}
\vspace{-5pt}
{\caption{$psl(2|2)$ root system.}  \label{fig A11}}
\vspace{-5pt}
\end{figure}
%
The root system of the $psl(2|2) = sl(2|2)/\mathbb{C} I \approx A(1,1)$
Lie superalgebra is shown on Figure~\ref{fig A11}.
The metric in the $\alpha_3$ direction of the root space is negative.
The even part of the algebra is just $sl(2) \oplus sl(2)$, leading to the
2 pairs of even roots: $\theta, -\theta; \alpha_3, -\alpha_3$. There are
8 odd roots: $\alpha_1,\alpha_2,\alpha_4,\alpha_5$ and their opposites,
all of them are isotropic. The $A(n,n)$
type Lie superalgebras have an interesting feature:
the number of simple roots is greater than the rank of the algebra,
in our case the number of simple roots is 3, and the rank is 2.
The roots $\alpha_1$ and $\alpha_2$ (as well as $\alpha_4$ and $\alpha_5$)
coincide since the action of Cartan generators on the
corresponding root elements of the Lie superalgebra
is identical, so the odd root spaces are two--dimensional.
To resolve the ``degeneracy'' of the odd roots
one should equip the root system with an additive ``charge'':
the ``charge'' of the even roots being zero, the ``charge''
of $\alpha_1$ and $\alpha_4$ being 1 and the ``charge''
of $\alpha_2$ and $\alpha_5$ being $-1$.

The set of simple roots for which $\theta$ is a highest root
is $\{ \alpha_1,\alpha_2,\alpha_3\}$. The products between
simple roots are $(\alpha_1|\alpha_1)=(\alpha_2 | \alpha_2 )=0$,
$(\alpha_3|\alpha_3)=-2$, $(\alpha_3|\alpha_1)=(\alpha_3|\alpha_2)=1$.
Other positive roots are obtained from the simple ones as
$\alpha_4=\alpha_1+\alpha_3$, $\alpha_5=\alpha_2+\alpha_3$,
$\theta=\alpha_1+\alpha_5=\alpha_2+\alpha_4$.

The W-algebra obtained by the quantum reduction of the minimal
gradation on $psl(2|2)$ is the $N=4$ superconformal algebra \cite{Ademollo:1976wv}.
This is a Lie
algebra generated by the Virasoro field $L$, three bosonic dimension--1
fields $J$, $J^+$ and $J^-$,
forming an $sl(2)$ affine vertex algebra on level $k_0=-k-1$,
and four fermionic dimension--$3/2$ fields
$G^+, G^-, \bar G^+, \bar G^-$
in two doublet representations
of the $sl(2)$:
\begin{equation}
\begin{aligned}
J^+ &\sim J^{\{\alpha_3\}}, \\
J&=J^{\{y\}},\\
J^-&\sim J^{\{{-\alpha_3}\}},
\end{aligned}
\qquad
\begin{aligned}
G^+ &\sim J^{\{-\alpha_1\}}, \\
G^-&\sim J^{\{{-\alpha_4}\}},
\end{aligned}
\qquad
\begin{aligned}
\bar G^+ &\sim J^{\{-\alpha_2\}}, \\
\bar G^-&\sim J^{\{{-\alpha_5}\}}.
\end{aligned}
\end{equation}
The central charge of the algebra is
\begin{equation}
c=-6(k+1).
\end{equation}
The \ope s of the algebra and the explicit reduction formulas can be found
in~\cite{KW1} (Section 8.4).

The twistings are parameterized by two numbers: $\e_1 \equiv \e_{\alpha_1}$
and $\e_2 \equiv \e_{\alpha_2}$. Other twistings are expressed as
$\e_{\alpha_3}=-\e_1-\e_2$, $\e_{\alpha_4}=-\e_2$, $\e_{\alpha_5}=-\e_1$.
The NS sector is given by $\e_1=\e_2=0$, the Ramond sector corresponds to
$\e_1=\e_2=1/2$. There is a $U(1)$ flow, which relates different sectors
\cite{Schwimmer:1986mf}, in particular NS and Ramond sectors are isomorphic.

The dual Coxeter number of $psl(2|2)$ is $h^\lor=0$.
In our case $\Delta_0^+=\{\alpha_3\}$, $\Delta_{1/2}^+=\{\alpha_4,\alpha_5\}$,
therefore $\rho_0=\alpha_3/2$ and $\rho_{1/2}^\natural=-\alpha_4^\natural$.
Denoting by $q=\Lambda$ the $J_0$ eigenvalue we write down the determinant
formula of the $N=4$ superconformal algebra in the case of
NS ($\e =\e_1=\e_2=0$)
and Ramond ($\e =\e_1=\e_2=1/2$) sectors:
\begin{equation}                \label{det A11}
\begin{aligned}
&\text{det}_{\widehat \eta}(k,q,h)=
k^
{\sum_{m,n \in \mathbb{N}} P_W (\widehat{\eta}-(0,m n))}
\!\prod_{m,n \in \mathbb{N}} \! {\mathcal{N}}_{n,m}^{\ \theta \, }(k,q,h)^
{P_W (\widehat{\eta}-(0, mn))}\\
& \times
\hspace{-5pt}
\prod_
{
m\in\frac{1}{2}+\e+\nz}
\hspace{-5pt}
{\mathcal N}_{1,m}^{\ \alpha_1}(k,q,h)^{2P_W^{(-1/2,m)}(\widehat{\eta}-(-1/2, m))}
\hspace{-5pt}
\prod_{
m\in\frac{1}{2}-\e+\nz}
\hspace{-5pt}
{\mathcal N}_{1,m}^{\ \alpha_4}(k,q,h)^{2P_W^{(1/2,m)}(\widehat{\eta}-(1/2, m))}
\\
& \times
\!
\prod_
{n \in \mathbb{N},\,
m\in\nz}
\!
{\mathcal N}_{n,m}^{\ \alpha_3}(k,q)^{P_W(\widehat{\eta}-n(1, m))}
\!
\prod_{
m,n \in \mathbb{N}}
\!
{\mathcal N}_{n,m}^{\, -\alpha_3}(k,q)^{P_W(\widehat{\eta}-n(-1, m))},
\end{aligned}
\end{equation}
where
\begin{align}
{\mathcal N}_{n,m}^{\ \theta}(k,q,h)&=h
-\frac{1}{4k}
\bigg(
(k m-n )^2-
4(q+1/2-\e)^2
\bigg)
+\frac{k+2}{4}-\e^2  ,
\\
{\mathcal N}_{1,m}^{ \alpha_1(\alpha_4)}
(k,q,h)&=
h
-m \big( k m \pm 2(q+1/2-\e) \big)
+\frac{k+2}{4}-\e^2,
\\
{\mathcal N}_{n,m}^{\pm \alpha_3}(k,q)&=
\mp 2(q+1/2-\e)+k m+n .
\end{align}

The partition generating functions are
\begin{align}
&\prod_{l=1}^\infty \frac{(1+x^{1/2} y^{l-1/2-\e})^2(1+x^{-1/2} y^{l-1/2+\e})^2}
{(1-y^l)^2 (1-x y^{l-1})(1-x^{-1} y^{l})}
=
\sum P_W(n_1,n_2)\, x^{n_1} y^{n_2}  ,\\
&
\frac{1}{1+x^{j_1} y^{j_2}}
\prod_{l=1}^\infty \frac{(1+x^{1/2} y^{l-1/2-\e})^2(1+x^{-1/2} y^{l-1/2+\e})^2}
{(1-y^l)^2 (1-x y^{l-1})(1-x^{-1} y^{l})}
=  \nonumber\\
&\phantom{aaaaaaaaaaaaaaaaaaaaaaaaaaaaaaaaaaaaaaaa} 
=\sum P_W^{(j_1,j_2)}(n_1,n_2)\, x^{n_1} y^{n_2}  .
\end{align}
The determinant formulae were conjectured in \cite{Kent:1987jy}.


\subsection{$G_2$}


\begin{figure}[h]
\vspace{-5pt}
\centering
\includegraphics[width=150pt]{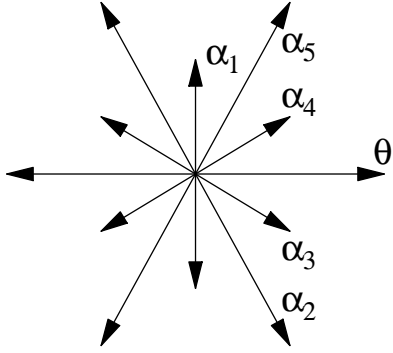}
\vspace{-5pt}
{\caption{$G_2$ root system.}  \label{fig G2}}
\vspace{-5pt}
\end{figure}
%
This is a simple exceptional Lie algebra with 14 generators.
The root system is shown on Figure~\ref{fig G2}.
$G_2$ is the only simple Lie algebra with a root square ratio equal
to 3. There are 6 long and 6 short roots. The simple roots are
$\alpha_1$ and $\alpha_2$, the products between them are
$(\alpha_1|\alpha_1)=2/3, (\alpha_2|\alpha_2)=2, (\alpha_1|\alpha_2)=-1$.
Other positive roots are $\alpha_3=\alpha_2+\alpha_1,
\alpha_4=\alpha_2+2\alpha_1, \alpha_5=\alpha_2+3\alpha_1,
\theta=2\alpha_2+3\alpha_1$.

The W-algebra obtained by the quantum reduction procedure
from the minimal gradation
on $G_2$ is generated by the Virasoro field, three dimension--1 bosonic
fields forming the $sl(2)$ affine vertex algebra on level $k_0=3k+5$ and
four bosonic dimension--$3/2$ fields in the quadruplet of the $sl(2)$:
\begin{equation}
\begin{aligned}
J^+ &\sim J^{\{\alpha_1\}}, \\
J&=J^{\{\sqrt{3} y\}},\\
J^-&\sim J^{\{{-\alpha_1}\}},
\end{aligned}
\qquad
\begin{aligned}
G^{++} &\sim J^{\{-\alpha_2\}}, \\
G^+&\sim J^{\{{-\alpha_3}\}},\\
 G^- &\sim J^{\{-\alpha_4\}}, \\
 G^{--}&\sim J^{\{{-\alpha_5}\}}.
\end{aligned}
\end{equation}
The central charge of the algebra is
\begin{equation}
c=\frac{14 k}{k+4}-6k\, .
\end{equation}

The twist numbers are parameterized by one discrete parameter
$\sigma$, which can take values $0$ and $1/2$, and one continuous
parameter $\e$: $\e_{\alpha_1}=2\e, \e_{\alpha_2}=-3\e+\sigma,
\e_{\alpha_3}=-\e+\sigma, \e_{\alpha_4}=\e+\sigma, \e_{\alpha_5}=3\e+\sigma$.
One gets the NS sector when $\e=\sigma=0$, the Ramond sector is obtained
when $\e=0, \sigma=1/2$.

The dual Coxeter number of $G_2$ is $h^\lor=4$, The Weyl vectors of
$\Delta_0=\{\alpha_1,-\alpha_1\}$ and
$\Delta_{1/2}=\{\alpha_2,\alpha_3,\alpha_4,\alpha_5\}$
are $\rho_0=\alpha_1/2$ and $\rho_{1/2}=(\alpha_4+\alpha_5)/2$.
The determinant formula of the minimal $W_k(G_2)$ algebra
in the NS and Ramond sectors becomes
\begin{equation}                \label{det G2}
\begin{aligned}
&\text{det}_{\widehat \eta}(k,q,h)=
(k+4)^
{\sum_{m,n \in \mathbb{N}} P_W (\widehat{\eta}-(0,m n))}
\!\prod_{m,n \in \mathbb{N}} \! {\mathcal{N}}_{n,m}^{\ \theta \, }(k,q,h)^
{P_W (\widehat{\eta}-(0, mn))}\\
& \times
\hspace{-5pt}
\prod_
{n\in \mathbb{N},
\atop
m\in\frac{1}{2}+\sigma+\nz}
\hspace{-5pt}
{\mathcal N}_{n,m}^{\ \alpha_2}(k,q,h)^{P_W(\widehat{\eta}-n(-3/2, m))}
\hspace{-5pt}
\prod_{
n\in \mathbb{N},
\atop
m\in\frac{1}{2}-\sigma+\nz}
\hspace{-5pt}
{\mathcal N}_{n,m}^{\ \alpha_5}(k,q,h)^{P_W(\widehat{\eta}-n(3/2, m))}
\\
& \times
\hspace{-5pt}
\prod_
{n\in \mathbb{N},
\atop
m\in\frac{1}{2}+\sigma+\nz}
\hspace{-5pt}
{\mathcal N}_{n,m}^{\ \alpha_3}(k,q,h)^{P_W(\widehat{\eta}-n(-1/2, m))}
\hspace{-5pt}
\prod_{
n\in \mathbb{N},
\atop
m\in\frac{1}{2}-\sigma+\nz}
\hspace{-5pt}
{\mathcal N}_{n,m}^{\ \alpha_4}(k,q,h)^{P_W(\widehat{\eta}-n(1/2, m))}
\\
& \times
\!
\prod_
{n \in \mathbb{N},\,
m\in\nz}
\!
{\mathcal N}_{n,m}^{\ \alpha_1}(k,q)^{P_W(\widehat{\eta}-n(1, m))}
\!
\prod_{
m,n \in \mathbb{N}}
\!
{\mathcal N}_{n,m}^{\, -\alpha_1}(k,q)^{P_W(\widehat{\eta}-n(-1, m))},
\end{aligned}
\end{equation}
where
\begin{align}
{\mathcal N}_{n,m}^{\ \theta}(k,q,h)&=h
-\frac{1}{4(k+4)}
\bigg(
\big(m(k+4)-n \big)^2+
 \nonumber \\
&  \phantom{aaaaaaaaaaaaaaa\,\,}
+\frac{4}{3}(q+2\sigma)(q+2\sigma+1)-(k+1)^2
\bigg)+\sigma^2,
\\
{\mathcal N}_{n,m}^{ \alpha_{2}(\alpha_5)}
(k,q,h)&=
h
-\frac{1}{4(k+4)}
\bigg(
4\big(m(k+4)-n \mp (q+2\sigma+1/2) \big)^2+ \nonumber \\
& \phantom{aaaaaaaaaaaaaaa\,\,}
+\frac{4}{3}(q+2\sigma)(q+2\sigma+1)-(k+1)^2
\bigg)+\sigma^2
 ,\\
 {\mathcal N}_{n,m}^{ \alpha_{3}(\alpha_4)}
(k,q,h)&=
h
-\frac{1}{4(k+4)}
\bigg(
4\big(m(k+4)-\frac{1}{3}n \mp \frac{1}{3}(q+2\sigma+1/2) \big)^2+ \nonumber \\
& \phantom{aaaaaaaaaaaaaaa\,\,}
+\frac{4}{3}(q+2\sigma)(q+2\sigma+1)-(k+1)^2
\bigg)+\sigma^2
 ,\\
{\mathcal N}_{n,m}^{\pm \alpha_1}(k,q)&=
\pm 2/3(q+2\sigma+1/2)+m(k+4)-n/3\, ,
\end{align}
here $q=\sqrt{3}\Lambda$ is the $J_0$ eigenvalue and $\sigma=0$ or $1/2$.
The partition function is
\begin{equation}
\begin{split}
\prod_{l=1}^\infty
\frac{1}
{(1-x^{3/2} y^{l-1/2-\sigma})(1-x^{1/2} y^{l-1/2-\sigma})
 (1-x^{-1/2} y^{l-1/2+\sigma})(1-x^{-3/2} y^{l-1/2+\sigma})} \\
\times \frac{1}
{(1-x\,y^{l-1})(1-y^l)^2(1-x^{-1}\,y^{l})}
=
\sum P_W(n_1,n_2)\, x^{n_1} y^{n_2} .
\end{split}
\end{equation}


\subsection{$osp(1|4)$}


\begin{figure}[h]
\vspace{-5pt}
\centering
\includegraphics[width=150pt]{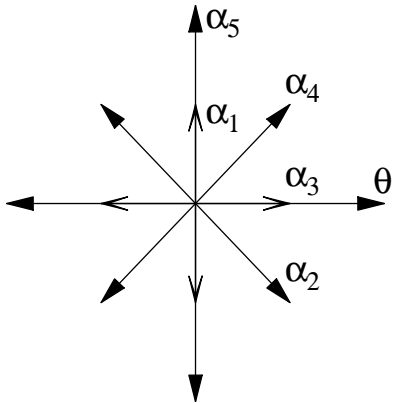}
\vspace{-5pt}
{\caption{$osp(1|4)$ root system.}  \label{fig osp14}}
\vspace{-5pt}
\end{figure}
%
The $osp(1|4) \approx B(0,2)$ Lie superalgebra has 14 generators:
10 even and 4 odd. The even subalgebra is the $so(5)$
Lie algebra. The root system is shown on Figure~\ref{fig osp14}.
There are 8 even roots and 4 odd roots.
The metric in both directions of the root space is positive.
The simple roots are $\alpha_1$ and $\alpha_2$. The defining
products are $(\alpha_1|\alpha_1)=1/2$, $(\alpha_2|\alpha_2)=1$
and $(\alpha_1|\alpha_2)=-1/2$. Other positive roots are given by
$\alpha_3=\alpha_1+\alpha_2$, $\alpha_4=2\alpha_1+\alpha_2$,
$\alpha_5=2\alpha_1$, $\theta=2\alpha_3$.

The minimal W-algebra obtained from the quantum reduction
of $osp(1|4)$ is generated along with the Virasoro field
by five dimension--1 fields which form an $osp(1|2)$ affine vertex algebra
on level $k_0=k+1$ and three dimension--$3/2$ fields in the triplet
representation of the $osp(1|2)$:
\begin{equation}
\begin{aligned}
J^+ &\sim J^{\{\alpha_5\}}, \\
j^+ &\sim J^{\{\alpha_1\}}, \\
J&=J^{\{ y\}},\\
j^-&\sim J^{\{{-\alpha_1}\}},\\
J^-&\sim J^{\{{-\alpha_5}\}},
\end{aligned}
\qquad
\begin{aligned}
G^+&\sim J^{\{{-\alpha_2}\}},\\
 G &\sim J^{\{-\alpha_3\}}, \\
 G^-&\sim J^{\{{-\alpha_4}\}},
\end{aligned}
\end{equation}
$j^+, j^-, G$ are fermionic fields, the rest are bosonic.
The central charge of the algebra is
\begin{equation}
c=\frac{6 k}{k+5/2}-6k-3/2 .
\end{equation}

There is a 2 parameter family of twistings: $\e \equiv \e_{\alpha_1}$
is continuous parameter, $\sigma=\e_{\alpha_3}$ is discrete, can take value
0 and $1/2$. Other twistings are expressed as $\e_{\alpha_2}=\sigma-\e$,
$\e_{\alpha_4}=\sigma+\e$, $\e_{\alpha_5}=2\e$. The NS sector corresponds
to $\e=\sigma=0$, the Ramond sector is obtained when $\e=0, \sigma=1/2$.

Inserting the values of the dual Coxeter number ($h^\lor=5/2$),
of the Weyl vectors for $\Delta_0^+=\{\alpha_1,\alpha_5\}$
($\rho_0=\alpha_5/4$) and for $\Delta_{1/2}^+=\{\alpha_4\}$
($\rho_{1/2}=\alpha_4/2$) into our general
minimal W-algebra determinant formula in
Section~\ref{Determinant formulae for minimal W-algebras}
one gets the determinant formula for the NS and Ramond sectors of the
minimal $W_k(osp(1|4))$ algebra:
\begin{equation}
\begin{aligned}
&\text{det}_{\widehat \eta}(k,q,h)=
(k+\tfrac{5}{2})^
{\sum_{m,n \in \mathbb{N}} P_W (\widehat{\eta}-(0,m n))}
\!\prod_{\scriptstyle m,n \in
\mathbb{N}, \atop \scriptstyle m-n \in 2\mathbb{Z}+2\sigma}
\! \nu_{n,m}(k,h)^ {P_W(\widehat \eta- (0,\frac{m n}{2}))}
\\
& \times
\hspace{-5pt}
\prod_
{n \in \mathbb{N},
\atop
m\in\frac{1}{2}-\sigma+\nz}
\hspace{-5pt}
{\mathcal N}_{n,m}^{\, \alpha_4}(k,q,h)^{P_W(\widehat{\eta}-n(1/2, m))}
\hspace{-5pt}
\prod_{
n \in \mathbb{N},
\atop
m\in\frac{1}{2}+\sigma+\nz}
\hspace{-5pt}
{\mathcal N}_{n,m}^{\, \alpha_2}(k,q,h)^{P_W(\widehat{\eta}-n(-1/2, m))}
\\
& \times
\!
\prod_
{n \in \mathbb{N},\,
m\in\nz,
\atop
m-n\in 2\mathbb{Z}+1}
\!
\nu_{n,m}^{\, \alpha_5}(k,q)^{P_W(\widehat{\eta}-(\frac{n}{2},\frac{m n}{2}))}
\!
\prod_
{m,n \in \mathbb{N},
\atop
m-n\in 2\mathbb{Z}+1}
\!
\nu_{n,m}^{-\alpha_5}(k,q)^{P_W(\widehat{\eta}-(-\frac{n}{2},\frac{m n}{2}))}
,
\end{aligned}
\end{equation}
where
\begin{align}
\nu_{n,m}(k,q,h)&=h
-\frac{1}{4(k+\frac{5}{2})}
\bigg(
\big(m(k+\frac{5}{2})-\frac{n}{2}\big)^2+ \nonumber \\
&\phantom{aaaaaaaaaaaa}
+(2q+\sigma)(2q+\sigma+1)-(k+1)^2
\bigg)+\frac{\sigma^2}{4},
\\
{\mathcal N}_{n,m}^{ \alpha_{4}(\alpha_2)}
(k,q,h)&=
h-
\frac{1}{4(k+\frac{5}{2})}
\bigg(
\big(
2m(k+\frac{5}{2})-n\pm (2q+\sigma+\frac{1}{2})
\big)^2+
\nonumber \\
&\phantom{aaaaaaaaaaaa}
+(2q+\sigma)(2q+\sigma+1)-(k+1)^2
\bigg)
+\frac{\sigma^2}{4},\\
\nu_{n,m}^{\pm \alpha_5}(k,q)&=
\pm (2q+\sigma+1/2)+m(k+5/2)-n/2\, ,
\end{align}
here $q=\Lambda$ is the eigenvalue of $J_0$, and $\sigma=0$
for the NS sector and $\sigma=1/2$ for the Ramond sector.
The partition generating function is given by
\begin{equation}
\begin{split}
\prod_{l=1}^\infty \frac{(1+x^{1/2} y^{l-1})(1+ y^{l-1/2+\sigma})(1+x^{-1/2} y^{l})}
{(1-x\, y^{l-1})(1-x^{1/2}y^{l-1/2-\sigma})(1-y^l)^2
(1-x^{-1/2} y^{l-1/2+\sigma})(1-x^{-1} y^{l})
}
=&\\
=
\sum P_W(n_1,n_2)\, x^{n_1} y^{n_2}  .&
\end{split}
\end{equation}


\section{Discussion}

\label{Discussion}

\setcounter{equation}{0}


We studied the quantum reduction of affine superalgebras in the
twisted case. This is also a subject of paper \cite{KW2}.
The methods and the results obtained are essentially the same.
However some details and the presentation are different.
The main difference is in the choice
of the triangular decomposition of the twisted loop algebra.
(Compare (\ref{triangular decomposition}) with (2.6--2.9) of \cite{KW2}.)
Also the different normal ordered product prescriptions are used
(see Appendix~\ref{Normal ordered product conventions}).
We consider only the case when
the Cartan subalgebra is untwisted ($\e(\fh)=0$),
the discussion in \cite{KW2} applies to the more general twisting
than one discussed here: the case
$\e(h) \ne 0$ for some $h\in \fh$ is also allowed.

We would like to show that our main result, the determinant
for the Ramond sector of minimal W-algebras is the same as in
\cite{KW2}.
Take the determinant formula of Kac and Wakimoto in \cite{KW2}, Theorem 4.2.
The Ramond sector corresponds to Example 4.1(b) in \cite{KW2}.
Using the values of $s_\alpha$ from this Example one can evaluate
the determinant factors (4.8--4.10) of \cite{KW2}.
Then it is easy to see that the first type factor ($\alpha(x)=0$)
coincides with our first type factor (\ref{Ramond a(x)=0}).
The other two factors are different by an expression proportional to
\begin{equation}             \label{statmnt}
  R=4(\rho_{1/2}^\natural | \rho_{1/2}^\natural+\rho_0)
-\frac{3}{8}\,\sigma-\frac{1}{2}\,
h^\lor(h^\lor-2),
\end{equation}
where $\sigma=1$, if $\theta/2 \in \Delta$ and $\sigma=0$
otherwise, $\rho_{1/2}$ and $\rho_0$ are defined
in~(\ref{rho 1/2}) and (\ref{rho 0}) respectively.
We prove here that $R=0$ for any simple Lie superalgebra $\fg$.
The proof is based
on the fact that the square of the Weyl vector
$(\rho|\rho)$ does not depend on the
choice of positive roots. We calculate it first for
the original choice of positive roots
$\Delta_+=\Delta_0^+ \cup
\Delta_{1/2} \cup
\{\theta\}$:
\begin{equation}
(\rho|\rho)=(\rho_0|\rho_0)
+\frac{1}{2}\left(\frac{1}{2}\, \text{sdim}\fg_{1/2}+1\right)^2.
\end{equation}
Now we define another set of positive roots ${\bar \Delta_+}$
by ``flipping''
the roots from $\Delta_{1/2}^-$ to the opposite ones:
$
{\bar \Delta_+}=\Delta_0^+ \cup
\Delta_{1/2}^+ \cup \Delta_{-1/2}^+
\cup \{\theta, \sigma \frac{\theta}{2}\}.
$
This set is ``generated'' by the element $h_0+t \, x$,
where $h_0 \in \fh^\natural$ is the Cartan element used
to split $\Delta_0$ and $\Delta_{1/2}$ to positive and negative
parts (see (\ref{Delta splitting})),
and $t$ is a sufficiently small positive number.
Now the new ${\bar \rho}$ is defined with respect
to ${\bar \Delta_+}$, and its square
\begin{equation}
({\bar \rho}|{\bar \rho})=
(\rho_0+2\rho_{1/2}^\natural|\rho_0+2\rho_{1/2}^\natural)
+\frac{1}{2}\left(-\frac{1}{2}\,\sigma +1\right)^2.
\end{equation}
One can check that $R=({\bar \rho}|{\bar \rho})-
(\rho|\rho)$.
So we proved that $R=0$ and therefore the determinant factors coincide.

The only factor which is missing in our determinant formula
comparing to the formula in \cite{KW2} is
$\varphi_0$, which is present only if $\theta/2 \in \Delta$.
This factor is a contribution of the $G^{\{{-\theta/2}\}}_0$
zero mode. But since (unlike \cite{KW2}) we let this operator act diagonally
on the highest weight vector (see (\ref{G0 hw})), we do not have this
factor.

The factor multiplicities are given by partition functions defined
with respect to $\Delta_W^+$ ,the set of positive roots of the
minimal W-algebra. The degrees are the same in the
present paper and in \cite{KW2}. (Again up to a small difference
in the case when there is a root $\theta/2$:
unlike our definition (\ref{Delta_W^+}), in \cite{KW2} an odd root $(0,0)$
is included in the set of positive W-algebra roots.)


\subsection*{Acknowledgment}

The author is grateful to Maria Gorelik for many fruitful
discussions and continuous support during the work on this
paper. The author also would like to thank Victor Kac for
correspondence and Anthony Joseph for comments on the first
version of the paper.
The results of this paper were presented at
the Dublin Institute for Advanced Studies in April 2004
and on the ``Algebraic Geometry and Representation Theory Seminar''
at the Weizmann institute of Science in May 2004,
the author wishes to thank these institutions
for the opportunity to give a talk.


\appendix


\section{Normal ordered product conventions}

\label{Normal ordered product conventions}

\setcounter{equation}{0}


In this Appendix we fix the normal ordering conventions.
Start from the \ope\ of two fields:
\begin{equation}
A(z)B(w)=\sum_{l \in N(A,B)-  \nz}
\frac{[AB]^{(l)} (w)}{(z-w)^l},
\end{equation}
where $N(A,B) \in \mathbb{Z}$ is the order of maximal singularity
in the \ope\ of $A$ and $B$.
In all the formulas in this paper the normal ordering sign
$\NO{\,}{\,}$ stands for the so called point splitting
normal ordering, widely used in a physical literature.
It is just the \ope\ with
singular terms removed:
\begin{equation}
  \NO{A(z)}{B(w)}=\sum_{l \in  \nz}
[AB]^{(-l)} (w)(z-w)^l,
\end{equation}
and then $\NO{A}{B}(w)=\NO{A(w)}{B(w)}$ is just the zero
order term in the operator product expansion of fields
$A(z)$ and $B(w)$:
\begin{equation}
\NO{A}{B}=[AB]^{(0)}.
\end{equation}
In our formalism the normal ordering is affected
by local properties of the fields only (when $z$ is close to $w$).
Global properties such as boundary conditions do not influence
the normal ordered product.

The normal ordered product is not associative in general,
$\NO{(\NO{A}{B})}{C} \ne \NO{A}{(\NO{B}{C})}$.
However if the fields are free, i.e.~the singular part
of their mutual \ope s include the identity field only
(e.g.~superghosts and superfermions in this paper),
then the normal ordered product is associative.

We introduce mode expansions of the fields
\begin{equation}
\begin{aligned}
A(z)&=\sum_{n \in -\Delta(A)+\e(A)+\mathbb{Z}}
A_n \, z^{-n-\Delta(A)},
\\
B(w)&=\sum_{n \in -\Delta(B)+\e(B)+\mathbb{Z}}
B_n \, w^{-n-\Delta(B)},
\\
\NO{A}{B}(w)&=\hspace{-10pt}
\sum_{n \in -\Delta(A)-\Delta(B)+\e(A)+\e(B)+\mathbb{Z}}
\hspace{-10pt}
\NO{A}{B}_n \, w^{-n-\Delta(A)-\Delta(B)},
\end{aligned}
\end{equation}
where
$\Delta(A), \Delta(B), \Delta(\NO{A}{B})=\Delta(A)+ \Delta(B)$
are conformal dimensions of correspondent fields.
The twistings $\e(A),\e(B)\in \mathbb{R}/\mathbb{Z}$
depend on the boundary conditions.
The mode $\NO{A}{B}_n$ is expressed in terms of $A_n$ and $B_m$:
\begin{equation}\label{NOepsilon}
\begin{aligned}
\NO{A}{B}_n &=
- \sum_{l=1}^{N(A,B)} { \epsilon(A) \choose l} \, [AB]^{(l)}_n+\\
&+
\sum_{m \in -\Delta(A) + \epsilon(A)-\nz}
 A_m B_{n-m}+
(-1)^{p(A) p(B)}
 \sum_{m \in -\Delta(A) +1+\epsilon(A)+\nz}
 B_{n-m} A_m \, ,
 \end{aligned}
\end{equation}
$\e(A)\in \mathbb{R}$ can be any number consistent with
the algebraic structure of the theory, $p(A)$ and $p(B)$ are the field
parities: $p(A)=0$ if $A$ is even (bosonic) and $p(A)=1$ if $A$ is
odd (fermionic).
This formula 
is derived in Appendix~E
of~\cite{Noyvert:2002mc},
it also follows from the twisted Borcheds identity.
The formula is well known in the untwisted case ($\e(A)= 0$),
in the twisted case there are additional terms
(the first sum in (\ref{NOepsilon})) coming
from the singular part of the \ope .

We would like to stress that our definition of the normal
ordered product is different from one convenient in the mathematical
literature, which uses the separation of a field to
``positive'' and ``negative'' parts (see e.g.~\cite{Kac book} for details):
\begin{equation}
A(z)_-=\sum_
{n \ge -\Delta(A)+1}
A_n\, z^{-n-\Delta(A)},
\quad
A(z)_+=\sum_
{n <-\Delta(A)+1}
A_n\, z^{-n-\Delta(A)}.
\end{equation}
Then in this formalism the normal ordered product
${\times \atop \times} A(z)B(w){\times \atop \times}$
is defined as
\begin{equation}
\textstyle {\times \atop \times} A(z)B(w){\times \atop \times}=
A(z)_+ B(w)+(-1)^{p(A)p(B)}B(w)A(z)_- \, .
\end{equation}
It is easy to show that in the untwisted case the two definitions
coincide:
\begin{equation}
\textstyle {\times \atop \times} A(z)B(w){\times \atop \times}=
\NO{A(z)}{B(z)} \qquad
\text{(untwisted case)}.
\end{equation}
But they are in general different in the twisted case.

The advantage of the point--splitting formalism is that expressions in terms of
conformal fields (e.g.~(\ref{L g}), (\ref{L gh}) or (\ref{L ne}))
do not change when one changes the boundary conditions.



\end{document}